\documentclass{IEEEtran}
\pdfoutput=1
\usepackage{epsfig, multirow, array, cite}
\usepackage{amssymb, amsmath}
\usepackage{epstopdf,subfigure, color}
\allowdisplaybreaks

\ifCLASSINFOpdf 
\else
\fi
\begin{document}

\title{On-off Switched Interference Alignment for Diversity Multiplexing Tradeoff Improvement in the 2-User X-Network with Two Antennas
}

\author{\IEEEauthorblockN{Young-bin Kim, Myung Gil Kang, \textit{Member, IEEE}, and Wan Choi,
\textit{Senior Member, IEEE}}
\thanks{Parts of this paper have been submitted to IEEE International Conference on Communications (ICC), Shanghai, China, May, 2019 \cite{ICC2019}.}
\thanks{Manuscript received January 14, 2018; revised July 2, 2018 and November 2, 2018; accepted
November 11, 2018. The associate editor coordinating the review of this paper and approving it for publication was A. Zaidi.}
\thanks{This work was supported by the National Research Foundation of Korea through the Korean Government (MSIT) under Grant NRF- 2016R1A2B4012099. }
\thanks{ Y.-b. Kim and M. G. Kang were with School of Electrical Engineering, Korea Advanced Institute of Science and Technology (KAIST), and are now with KDDI Research, Inc., Saitama 356-8502, Japan and with Department of Information Systems and Techonology, Mid Sweden University, 851 70 Sundsvall, Sweden, respectively. }
\thanks{ W. Choi are with School of Electrical Engineering, Korea Advanced Institute of Science and Technology (KAIST), Daejeon 34141, Korea (e-mail: wchoi@kaist.edu).}
}
\maketitle
\newtheorem{lemma}{Lemma}
\newtheorem{theorem}{Theorem}
\newtheorem{definition}{Definition}
\newtheorem{corollary}{Corollary}
\newtheorem{remark}{Remark}

\vspace{-0.4in}
\begin{abstract} 
To improve diversity gain in an interference channel and hence to maximize diversity multiplexing tradeoff (DMT), we propose on-off switched interference alignment (IA) where IA is intermittently utilized by switching IA on/off.  For on-off switching, either IA with symbol extension or  IA with Alamouti coding is adopted in this paper.
Deriving and analyzing DMT of the proposed schemes, we reveal that the intermittent utilization of IA with simultaneous non-unique decoding can improve DMT in the 2-user X-channel with two antennas. Both the proposed schemes are shown to achieve 
diversity gain of 4 and DoF per user of $\frac{4}{3}$. 
In particular, the on-off switched IA with Alamouti coding,
to the best of our knowledge, surpasses any other existing schemes for the 2-user X-channel with two antennas and nearly approaches the ideal DMT.
\end{abstract}
\vspace{-0.1in}

\section{Introduction}
To deal with interfering signals and maximize the sum degrees of freedom (DoF),  many recent studies have paid attention to interference alignment (IA), also referred to multiplexing gain \cite{IAIC, IAX1, IAX2}. With IA, interfering signals can be aligned in minimal dimensions separated from the desired signals, and hence the desired signals can be decoded without interference.  It was shown that IA achieves the maximum DoF in interference channels \cite{IAIC} and in the 2-user
multiple input multiple output (MIMO) X-channel \cite{IAX1,IAX2}. In particular, in the
2-user MIMO X-channel with $M$ antennas at each node, it was reported that IA achieves the optimal sum DoF of $\frac{4M}{3}$.

As aforementioned, IA  was  proposed to improve DoF in various interference networks, and the IA studies have been quite mature in terms of optimal DoF,  as well as implementation issues \cite{R1}. However, diversity schemes guaranteeing optimal DoF in interference channels have not been sufficiently focused on, although it is a natural viewpoint shift from multiplexing to diversity in communications engineering.
In this context, contrary to the works that highlight achievable rate maximization in terms of sum DoF, recent works \cite{IND1, IND2, IAAla, IASTBC3, IASTBC4, IASTBCM}  focused on
improving diversity gain in interference channels. Linear transmission schemes were proposed in \cite{IND1, IND2} to achieve full diversity gain in the two X-channel equipped with arbitrary numbers of antennas at the transmitters and receivers. However, they
could not achieve the sum DoF that scales with the number of antennas at the transmitters and receivers.
In  \cite{IAAla, IASTBC3, IASTBC4, IASTBCM}, IA schemes combined with Alamouti code \cite{Ala}, space time block coding (STBC) \cite{STBC} or Srinath-Rajan STBC \cite{SRSTBC}, were developed. Specifically, in the 2-user X-channel
with two antennas at every node, the authors of \cite{IAAla} proposed an IA scheme that combines Alamouti coding and transmit beamforming over extended symbol times. It was shown that the proposed scheme achieves  maximal diversity gain of 2 and maximal sum DoF of $\frac{8}{3}$,  using only local channel state information at  transmitters (CSIT). This result was extended to the cases when the number of antennas
is 3 in \cite{IASTBC3} and 4 in \cite{IASTBC4},  using STBC with local CSIT. With the devised scheme, maximal diversity gains of 3 and 4 were shown to be achievable, respectively, while maximal sum DoF of $\frac{4\times3}{3}$ and $\frac{4\times4}{3}$ were achievable, respectively. Recently, the work was generalized  for an arbitrary number of antennas in \cite{IASTBCM}.

To understand the relationship between achievable diversity and multiplexing gains, 
diversity multiplexing tradeoff (DMT) \cite{dmt} has been popularly used  in various channels  \cite{DMTCZ, DMTICK, RecDMT1, RecDMT2}. 
 In \cite{DMTCZ} and \cite{DMTICK}, DMT was improved via the time-sharing scheme between IA and joint decoding in a 4-user clustered Z interference channel and $K$-user interference channel, respectively. The authors of \cite{RecDMT1} derived the DMT at the secondary receiver for the multiple-access channel and
user-selection schemes in an interweave multiuser cognitive radio
system,  considering the spectrum sensing effect.
In \cite{RecDMT2}, 
the authors analyzed DMT of a dynamic quantize-map-and-forward strategy in half-duplex single-relay networks. It showed that by optimizing listening time of the relay, the strategy can achieve the optimal DMT for half-duplex single-relay networks with local channel state information.

In the aforementioned works\cite{IAAla, IASTBC3, IASTBC4, IASTBCM}, the analyzed diversity gain and DoF correspond to the point $(d,0)$ and $(0,r)$ in the DMT
 domain. However, the DMT curve obtained by a linear function connecting the two points is not optimal in 2-user X-channel
with two antennas.  The sub-optimality of the linear DMT curve results from the fact that IA is continuously applied over the whole transmission time.
This observation poses a fundamental question: Can DMT be improved by intermittent usage of IA in interference channels? For answering it, we propose on-off switched IA which allows intermittent utilization of IA. It has been known that time-division multiplexing (TDM) of multiple independent schemes, which exploits  different codewords for each scheme, cannot improve DMT since the bottleneck among them determines DMT. However, interestingly, we reveal that the on-off switched beamformer with simultaneous non-unique decoder, which exploits a single codeword, can improve DMT compared to that of each single strategy (i.e., when the portion of on-off switching is 0 or 1). 
In the proposed on-off switched IA scheme, according to the portion of IA utilization, a part of the codeword (i.e., encoded message) is transferred with the given transmit and receive beamformers when the beamformer 
is switched \emph{on}. On the other hand, the remaining part is delivered without beamforming when the beamformer is switched \emph{off}. 
To the best of our knowledge, the DMT improvement by intermittent utilization of IA is justified for the first time in this paper.  

This paper ultimately aims to enhance diversity gain for all possibly achievable multiplexing gain regimes (i.e., for various data rate) in the given interference channel model. To this end,
we analyze DMT of two on-off switched IA schemes in the 2-user X-channel with two antennas, which optimally switches on/off either IA based on  symbol extension or  IA using Alamouti coding. At the receiver, the simultaneous non-unique decoder \cite{snd} is adopted. 
We derive DMT for the two proposed schemes in closed form and show that both schemes achieve maximal
diversity gain of 4 and maximal sum DoF of $\frac{8}{3}$, even with two antennas at each node. For each on-off switched IA scheme, the optimal time portion of  IA utilization is determined to achieve the highest diversity gain under given multiplexing gain.
If we optimally switch the IA scheme with symbol extension on/off, it outperforms the  IA scheme combined with Alamouti coding in  \cite{IAAla} for $0 \leq r \leq 1$, where $r$ is the multiplexing gain, although Alamouti coding is not used in the proposed scheme.
If on-off switching is applied to the IA scheme with Alamouti scheme and the portion of IA utilization is optimized,
to the best of our knowledge, it surpasses any other existing schemes in the 2-user X-channel with two antennas and each user achieves the DMT that nearly approaches $d(r)=4-3r$. We also discuss scalability of the proposed on-off switched IA in terms of the number of antennas at each node.

The rest of this paper is organized as follow. 
Section II describes the system model. 
The achievable DMT of the proposed schemes is analyzed in Section III. 
The time portion of IA utilization in the proposed schemes is optimized in Section IV. 
The extensions to more than two antennas at each node are discussed in Section V.
Finally, we draw conclusions in Section VI.

\section{System Model and Preliminaries}

\subsection{System Model}
\begin{figure}[!t]
\centering
  \includegraphics[width=0.8\columnwidth]{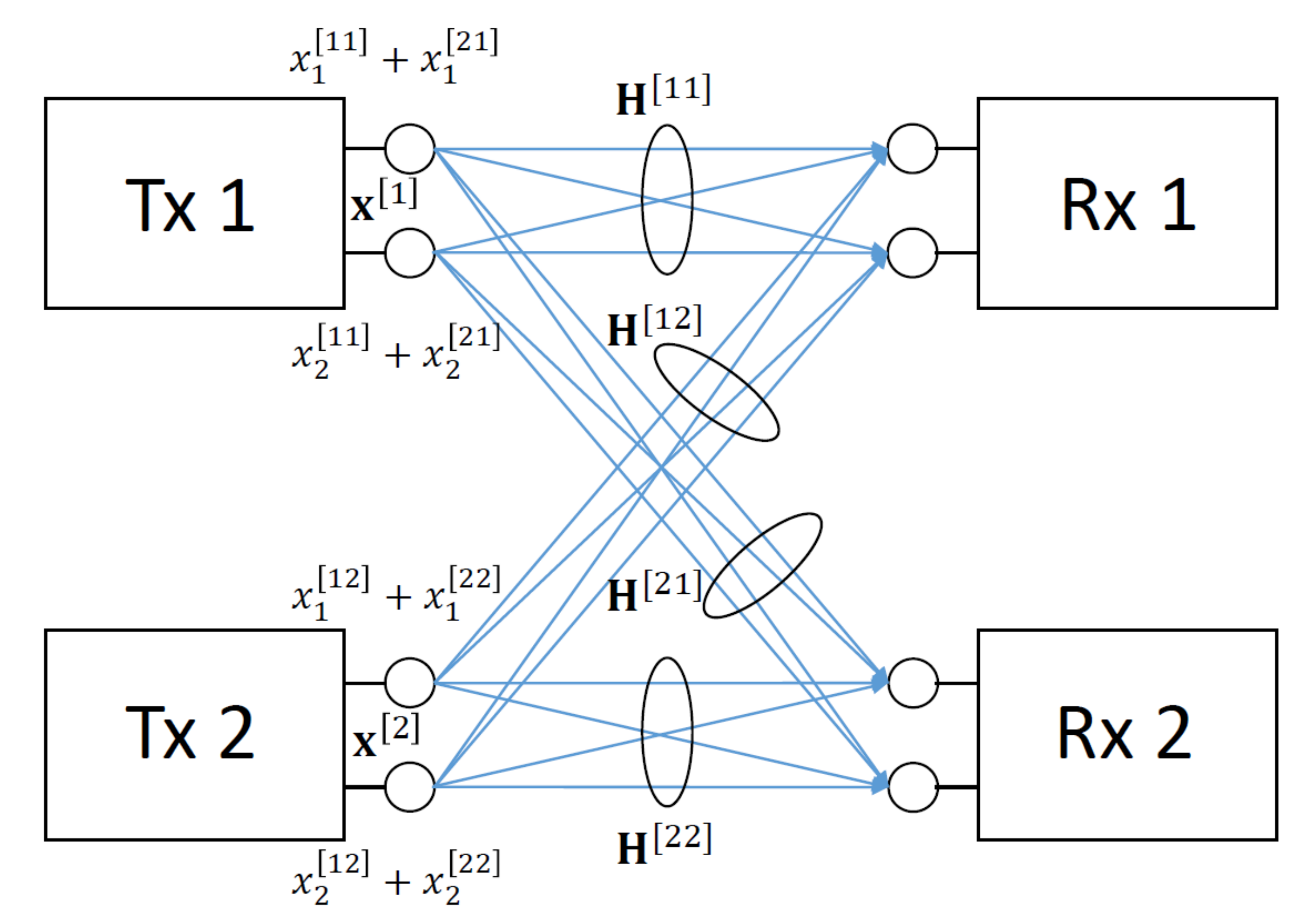}\\
  \caption{2-user X-channel with two antennas}
  \label{fig:P1}
\end{figure}

We consider a 2-user MIMO X-channel in wireless communication
as shown in Fig. \ref{fig:P1}. The system consists of
$2$ transmitters and $2$ receivers with two antennas each.
The received signal at receiver $i$ is given by
\begin{equation}
\label{Eq:eq1a} \mathbf{y}^{[i]}=
\sum_{j=1,2}\mathbf{H}^{[ij]}\mathbf{x}^{[j]}+\mathbf{z}^{[i]}
\end{equation}
where $\mathbf{y}^{[i]}=[y_1^{[i]}~~y_2^{[i]}]^T$, the
transmitted signal vector from transmitter $j$,
$\mathbf{x}^{[j]}=[x_1^{[1j]}+x_1^{[2j]}~~x_2^{[1j]}+x_2^{[2j]}]^T
\in \mathcal{C}^{2 \times 1}$ with a power constraint $P$, 
$y_k^{[i]}$ denotes the received signal with antenna $k$ at receiver $i$, and
$x_k^{[ij]}$ denotes the transmitted signal with antenna
$k$ from transmitter $j$ to receiver $i$.
$\mathbf{H}^{[ij]}\in \mathcal{C}^{2 \times 2}$ represents the
fading channel matrix from transmitter $j$ to receiver
$i$ and is given by
\begin{equation}
\mathbf{H}^{[ij]} = \left[
                  \begin{array}{cc}
                    h_{11}^{[ij]} & h_{12}^{[ij]} \\
                    h_{21}^{[ij]}  & h_{22}^{[ij]} \\
                  \end{array}
                \right]
\end{equation}
where $h_{BA}^{[ij]}$ denotes the channel coefficient from
antenna $A$ of transmitter $j$ to antenna $B$ of
receiver $i$ and
is assumed to be a complex Gaussian random variable with zero mean and unit variance $\sim
\mathcal{CN}(0,1)$. 
$\mathbf{z}^{[i]}=[z_1^{[i]}~~z_2^{[i]}]^T$ denotes the additive white Gaussian
noise (AWGN) vector, of which elements are independent and identically distributed complex Gaussian random variables with zero mean and variance  $N_0$
 $ \sim \mathcal{CN}(0,N_0)$.
 Channels between transmitters and receivers are assumed to be quasi-static.
Throughout our paper, we assume that each transmitter knows perfect CSI locally (i.e., local CSIT) and each receiver knows perfect
CSI globally (i.e., CSIR), as assumed in \cite{IAAla, IASTBC3, IASTBC4}. 

Each transmitter transmits an independent
packet consisting of a codeword obtained from random Gaussian
codebook. We assume fixed data rate for input as in \cite{IAAla, IASTBC3, IASTBC4, IASTBCM}
so that we exclude adaptive modulation/transmission according to channel conditions, which is not viable under local CSIT.

At the receiver, the simultaneous non-unique decoder \cite{snd} is adopted. 
The entire codeword is decoded by obeying the rule of simultaneous non-unique decoding.
Then, the transmitted message is recovered from the decoded codeword.

Signal-to-noise ratio (SNR) at each
receiver is denoted by $\rho$, i.e., $\rho=\frac{P}{N_0}$.
$\mathbb{R}^N$ is the set of real $N$-tuples, while
$\mathbb{R}^{N+}$ denotes the set of nonnegative real $N$-tuples.
For any set $\mathcal{O}\subset\mathbb{R}^N$, the intersection of
the set and $\mathbb{R}^{N+}$ is denoted by $\mathcal{O}^+$,
\emph{i.e.,} $\mathcal{O}^+=\mathcal{O}\cap\mathbb{R}^{N+}$.
Define $f(\rho)\doteq \rho^v$ if $\lim_{\rho\rightarrow\infty}\frac{\log(f(\rho))}{\log(\rho)}=v$.
$[a:b]$ denotes a set, $\{x \in \mathbb{R} | a\leq x \leq b \}$, where $\mathbb{R}$ is the set of real values.

\subsection{Preliminary: Diversity Multiplexing Tradeoff}
Key notations about DMT introduced in
\cite{dmt} can be defined as follows.
Assuming that $h$ is a Gaussian
random variable with zero mean and unit variance, the probability
density function (pdf) of the exponential order of $1/|h|^2$ can be given by
$p_v=\lim_{\rho\rightarrow\infty}\ln(\rho)\rho^{-v}\exp(-\rho^{-v})$,
where
$v=-\lim_{\rho\rightarrow\infty}\frac{\log(|h|^2)}{\log(\rho)}$. 
By taking the limit as $\rho$ goes to infinity, the pdf reveals that
\begin{align}
p_v\doteq \left\{
\begin{array}{ll}
\rho^{-\infty}=0, &\textrm{for } v< 0,\\
\rho^{-v}, &\textrm{for }v \geq 0.
\end{array}
\right.\label{eq:exporder}
\end{align}
Thus, for independent random variables $\{v_j\}_{j=1}^{K}$
distributed identically to $v$, the probability $\mathcal{P}_{\mathcal{O}}$
that $(v_1,\ldots,v_K)$ belongs to a set ${\mathcal{O}}$
 can be characterized by
\begin{align}
\label{eq:expresult}
P_{\mathcal{O}}\doteq \rho^{-d_0}, \quad\text{for }\quad d_0=\inf_{(v_1,\ldots,v_K)\in {\mathcal{O}}^+}\sum_{j=1}^{K}v_j
\end{align}
provided that ${\mathcal{O}}^+$ should be non-empty. Note that
multiplexing and diversity gains are defined, respectively, by
$r_j=\lim_{\rho\rightarrow \infty}\frac{ R_j(\rho)}{\log\rho} \quad \mathrm{and}\quad d_j=\lim_{\rho\rightarrow \infty}-\frac{\log\textrm{P}_{\textrm{out}, j}(\rho)}{\log\rho}$ where $R_j(\rho)$ represents the transmission rate and $\textrm{P}_{\textrm{out}, j}(\rho)$ denotes the outage probability
for receiver $j$.  For simplicity, we assume that the transmission rates for all users are the same, \emph{i.e.}, $r_j=r$ and $d_j=d$, $\forall j=1,\ldots, K$.
Note that the outage probability instead of the error probability can be used for DMT analysis since the dominant scales of the error probability and the outage probability are the same if the codeword length is sufficiently long \cite{dmt}. The optimal DMT of the point-to-point $m\times n$ MIMO channel, if the provided block length $l$ satisfies $l\geq m+n-1$ \cite{dmt}, is given by
\begin{align}
d^*_{m,n}(r)=(m-r)(n-r)
\end{align}
that is the piecewise-linear function connecting points $(r, d^*_{m,n}(r))$ for every integer $r \leq \min(m,n)$. For the MIMO multiple access channel (MAC) with $K$ transmitters, the outage probability can be used for DMT analysis if $l\geq Km+n-1$ \cite{MACdmt}. In this paper, we assume $l\geq 2K+1$ since we consider the two-antenna case.

\subsection{Preliminary: Simultaneous Non-unique Decoding}

\begin{figure}[!t]
\centering
  \includegraphics[width=1\columnwidth]{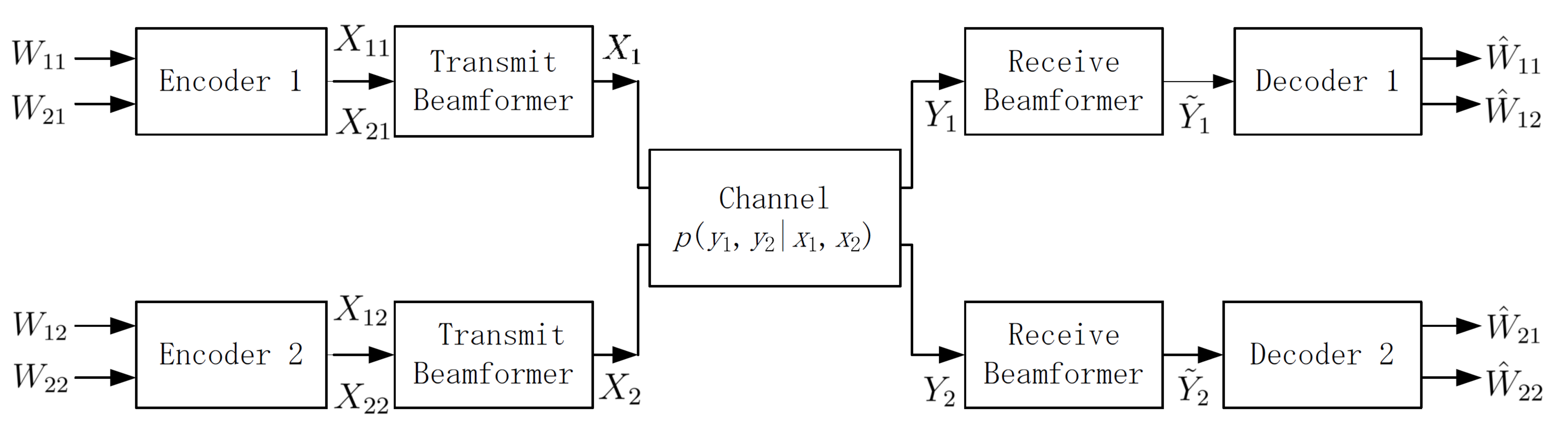}\\
  \caption{The block diagram of the simultaneous non-unique decoder in the 2-user-pair MIMO discrete memoryless X-channel}
  \label{fig:snd}
\end{figure}

We consider the simultaneous non-unique decoder at the receivers in MIMO X-channel~\cite{snd}.
It is shown that simultaneous non-unique decoding  with random codebook achieves the performance of optimal maximum likelihood decoding
in terms of capacity region. That is, the simultaneous non-unique decoder can attain the same performance of the combination of
simultaneous decoding (SD) and treating interference as noise (IAN). With the simultaneous non-unique decoder, the conceptual unification of
the aforementioned two methods of decoding with a single decoder is accomplished.

We characterize the rate region of the 2-user-pair MIMO discrete memoryless X-channel achieved by simultaneous non-unique decoding. As shown in Fig. \ref{fig:snd},
the 2-user-pair discrete memoryless X-channel is defined as $p(y_1, y_2|x_{1},x_{2})$ with input alphabets $\mathcal{X}_{1}$ and $\mathcal{X}_{2}$
and output alphabets $\mathcal{Y}_{1}$ and $\mathcal{Y}_{2}$.
Define a $(2^{nR_{11}}, 2^{nR_{21}}, 2^{nR_{12}}, 2^{nR_{22}}, n)$ code that is limited to randomly generated code ensemble with a special structure.
Define linear beamformer functions at transmitter and receiver as $l_T(\cdot)$ and $l_R(\cdot)$, respectively.
In Fig. \ref{fig:snd}, $l_T(x_{ij},x_{ij})=x_j$ and $l_R(y_k)=\tilde{y}_k$, $\forall i,j,k=1,2$.

In the discrete memoryless X-channel, 
it is shown in \cite{snd} that the rate region achievable by simultaneous non-unique decoding can be represented as
$\mathcal{R}=\mathcal{R}_1\cap\mathcal{R}_2$
where $\mathcal{R}_i$, $\forall i=1,2$ denotes the rate region of receiver $i$ and
$\mathcal{R}_i=\mathcal{R}_{i,\textrm{IAN}}\cup\mathcal{R}_{i,\textrm{SD}}$
where $\mathcal{R}_{i,\textrm{IAN}}$ and $\mathcal{R}_{i,\textrm{SD}}$ are the achievable rate regions of receiver $i$ by IAN and SD, respectively.

For receiver 1, $\mathcal{R}_{1,\textrm{IAN}}$ is the set of rate pairs $(R_1,R_2)$ such that
\begin{align}
R_{1,\textrm{IAN}}=R_{11,\textrm{IAN}}+R_{12,\textrm{IAN}}\leq I(X_{11},X_{12};\tilde{Y}_{1}).
\end{align}
Meanwhile,  $\mathcal{R}_{1,\textrm{SD}}$ is the set of rate pairs $(R_1,R_2)$ such that
\begin{align}
R_{1,\textrm{SD}}&=R_{11,\textrm{SD}}+R_{12,\textrm{SD}}\leq I(X_{11},X_{12};\tilde{Y}_{1}|X_{21},X_{22})\\
R_{2,\textrm{SD}}&=R_{21,\textrm{SD}}+R_{22,\textrm{SD}}\leq I(X_{21},X_{22};\tilde{Y}_{1}|X_{11},X_{12})\\
R_{1,\textrm{SD}}&+R_{2,\textrm{SD}}\leq I(X_{11},X_{12},X_{21},X_{22};\tilde{Y}_{1}).
\end{align}
For receiver 2, the achievable rate region can be obtained in a similar way to the receiver 1 case.

\section{DMT Analysis of the On-off Switched IA with Simultaneous Non-unique Decoding}

In this section, we explain how the on-off switched IA scheme operates and why it can improve DMT performance. To demonstrate the DMT improvement by our proposed scheme in the interference channel, we consider the two types of on-off switched IA schemes: On-off switching is applied to  either  IA with symbol extension or IA with Alamouti coding in the 2-user two-antenna X-channel. 
After we describe both IA schemes, we
derive DMT of the 2-user two-antenna X-channel when the schemes are used.

\begin{figure}[!t]
\centering
  \includegraphics[width=1\columnwidth]{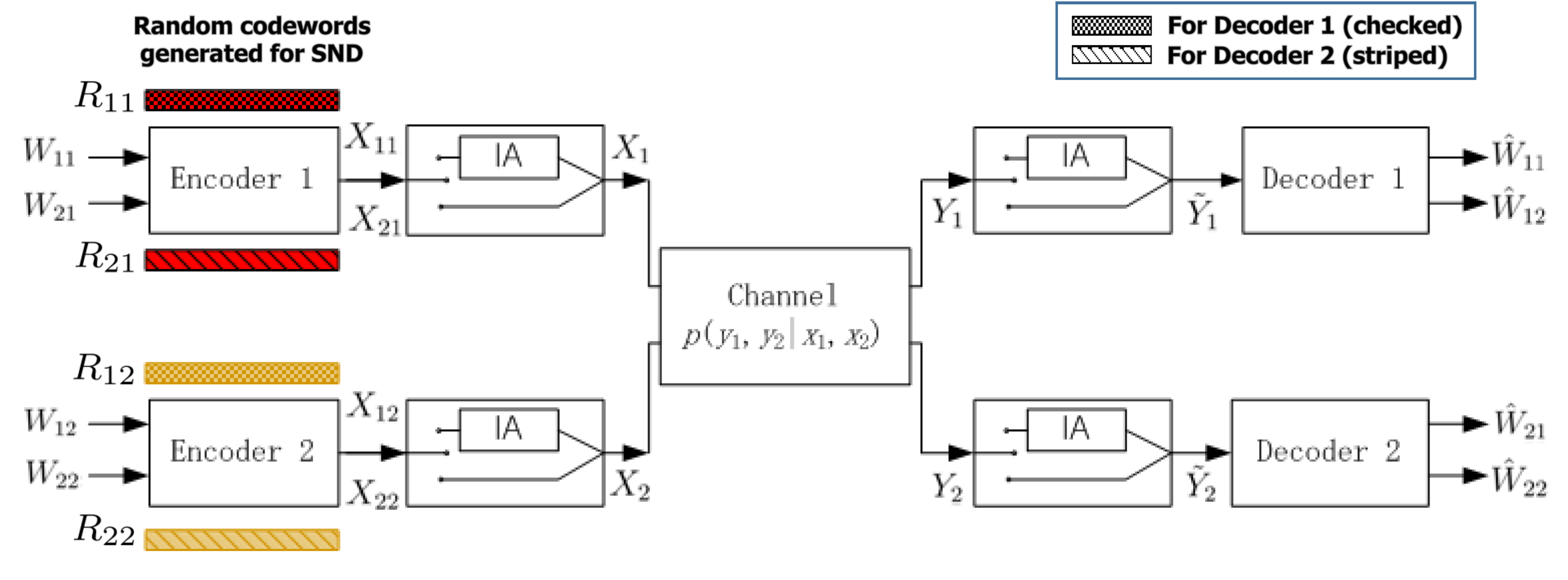}\\
  \caption{Encoding for the transmission in 2-user X-channel}
  \label{fig:onoff1}
\end{figure}
\begin{figure}[!t]
\centering
  \includegraphics[width=1\columnwidth]{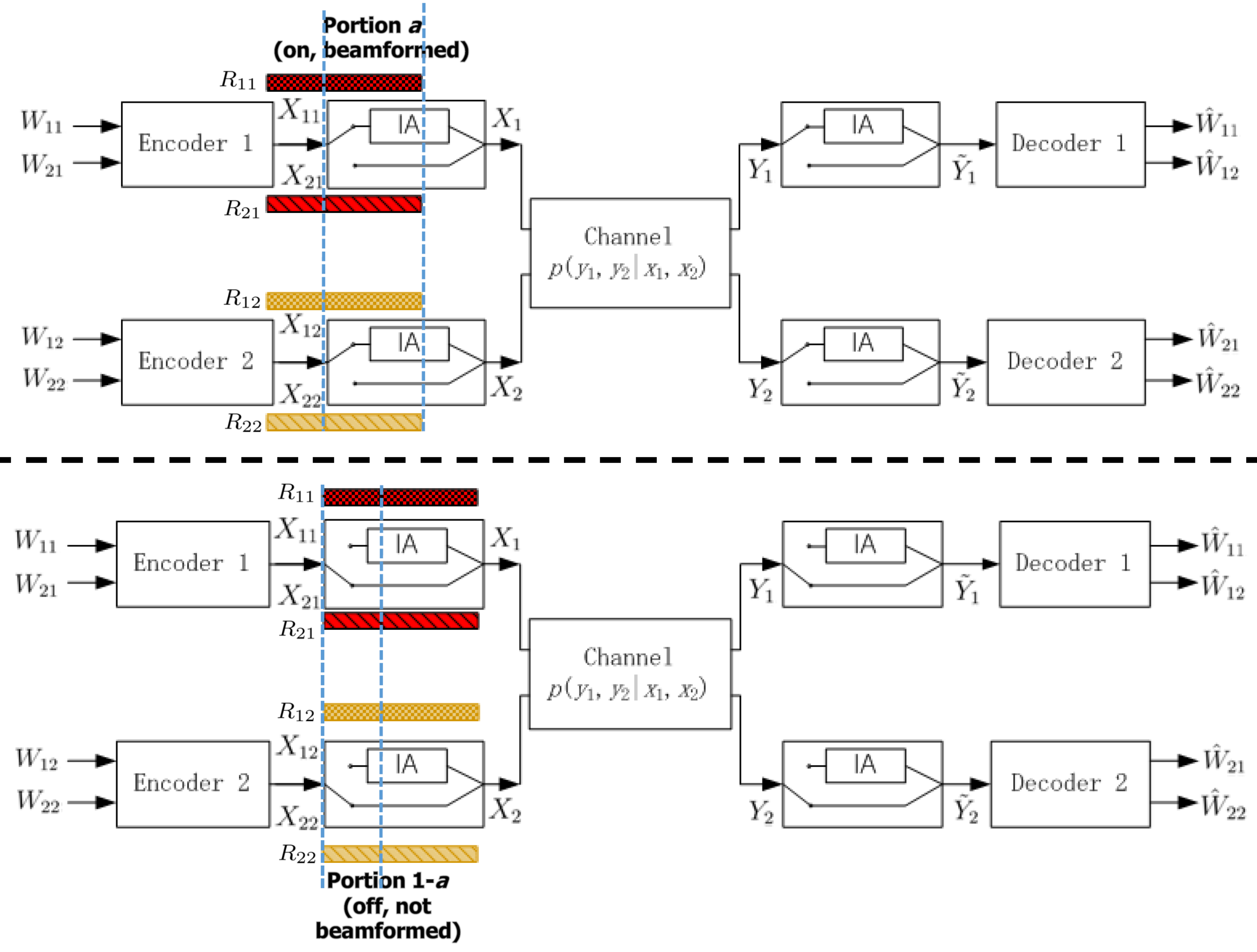}\\
  \caption{On-off switched beamforming at the transmitters in 2-user X-channel}
  \label{fig:onoff2}
\end{figure}

\subsection{On-off Switched IA with Simultaneous Non-unique Decoding}

\begin{figure}[!t]
\centering
  \includegraphics[width=1\columnwidth]{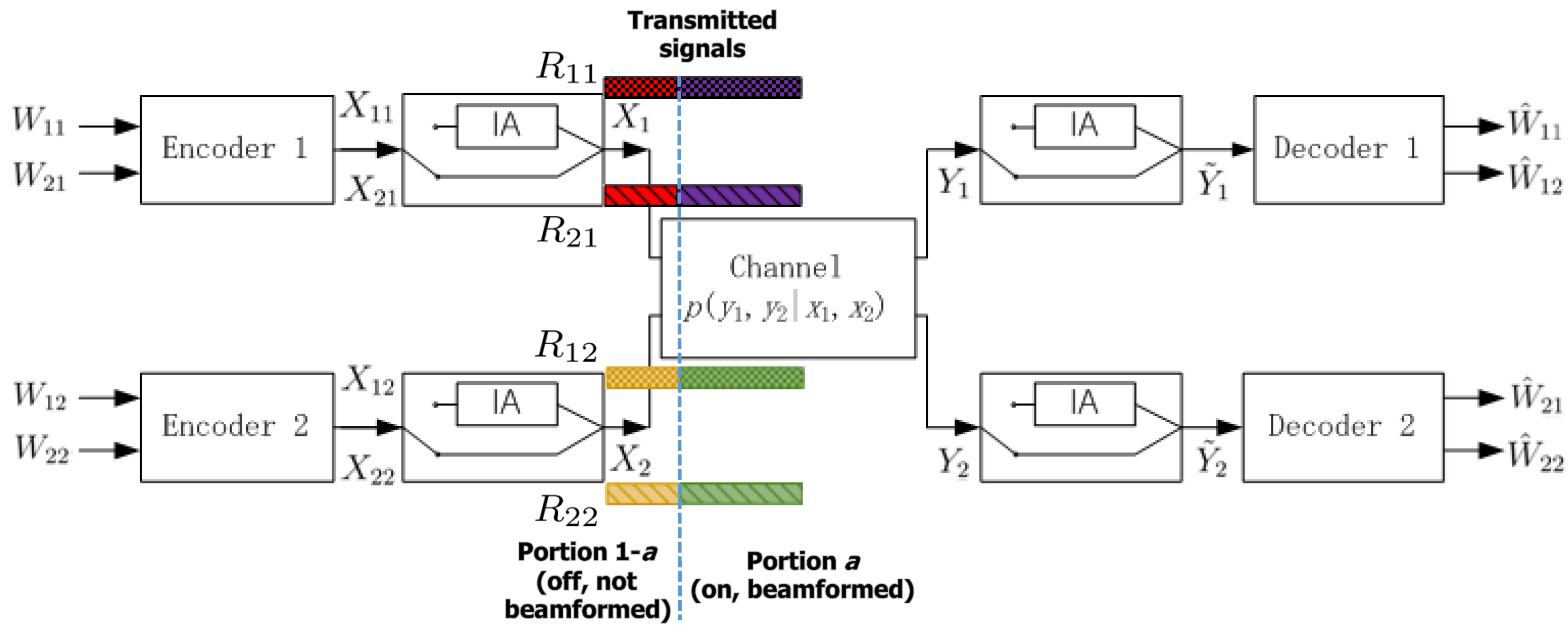}\\
  \caption{The transmitted signals after encoding and on-off switched IA in 2-user X-channel}
  \label{fig:onoff3}
\end{figure}

In order to improve DMT of various interference channels, especially the 2-user MIMO X-channel as one of examples, we propose the on-off switched IA.
For the proposed on-off switched IA, beamformer blocks at both transmitter and receiver consist of
the on-off switched beamformer that opportunistically switches on/off the pre-determined beamformer, as shown in Figs. \ref{fig:onoff1}-\ref{fig:onoff5}.
In this subsection, we describe how the on-off switched IA operates. 
Overall, there are five steps: (1) Encoding, (2) On-off switched beamforming (based on IA) at the transmitters, (3) Transmitting the beamformed signals, (4) On-off switched beamforming at the receivers, and (5) Simultaneous non-unique decoding. The details of each step are as follows.

\begin{enumerate}
\item[(1)] Encoding:
As shown in Fig. \ref{fig:onoff1}, for simultaneous non-unique decoding, input codewords are generated randomly. 
We basically assume that dark and light gray-shaded messages are encoded by transmitter 1 and 2, respectively. Let the checked and diagonally striped messages be intended to receiver 1 and 2, respectively. 
\begin{figure}[t]
\centering
  \includegraphics[width=1\columnwidth]{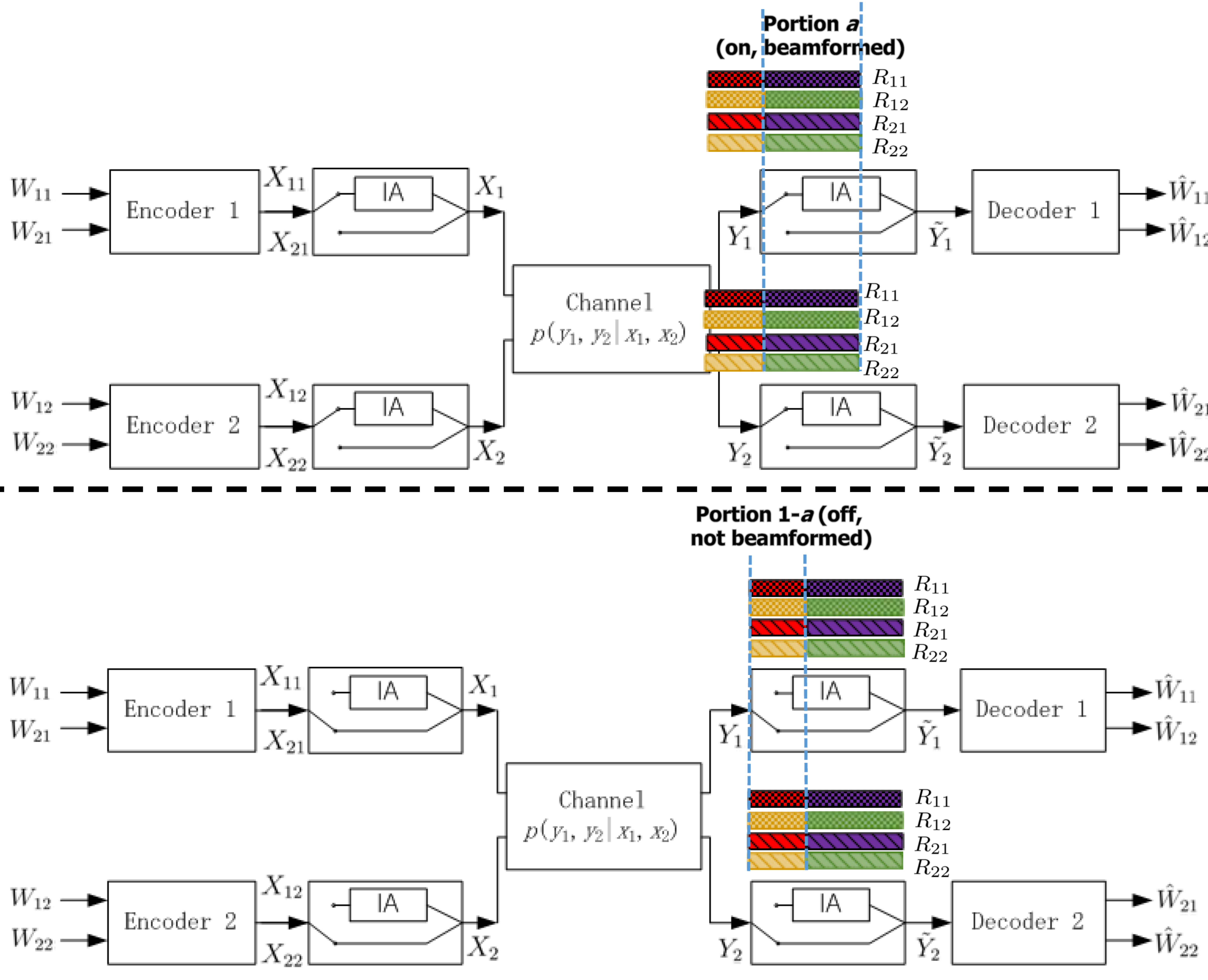}\\
  \caption{On-off switched IA at the receivers in 2-user X-channel}
  \label{fig:onoff4}
\end{figure}

\item[(2)]On-off switched beamforming at the transmitters:
After the codewords are encoded, the transmit beamformer is intermittently used by switching it on/off. 
As shown in Fig. \ref{fig:onoff2}, 
for a portion of $a$, each codeword is  precoded by the IA-based beamformer (dark and light gray-shaded rectangles next to the text `Beamformer' for transmitter 1 and 2, respectively), i.e., \emph{on}. Otherwise, for the remaining portion, $1-a$, each codeword is not precoded (not shaded rectangles), i.e., \emph{off}.
The criterion whether the beamformers at both the transmitter and receiver are on or off is based on
the pre-determined pattern of IA utilization, which enables to achieve the highest diversity gain for a given multiplexing gain.
The on-off pattern is assumed to be known at the both transmitter and receiver.
Suppose that the optimal portion of IA utilization is $a^*(r)$ according to a given multiplexing gain $r$. Then $a^*(r)$ of each codeword is beamformed while $1-a^*(r)$ of  each codeword is not beamformed. The optimization of the portion of IA utilization will be addressed in Section VI.

\begin{figure}[t]
\centering
  \includegraphics[width=1\columnwidth]{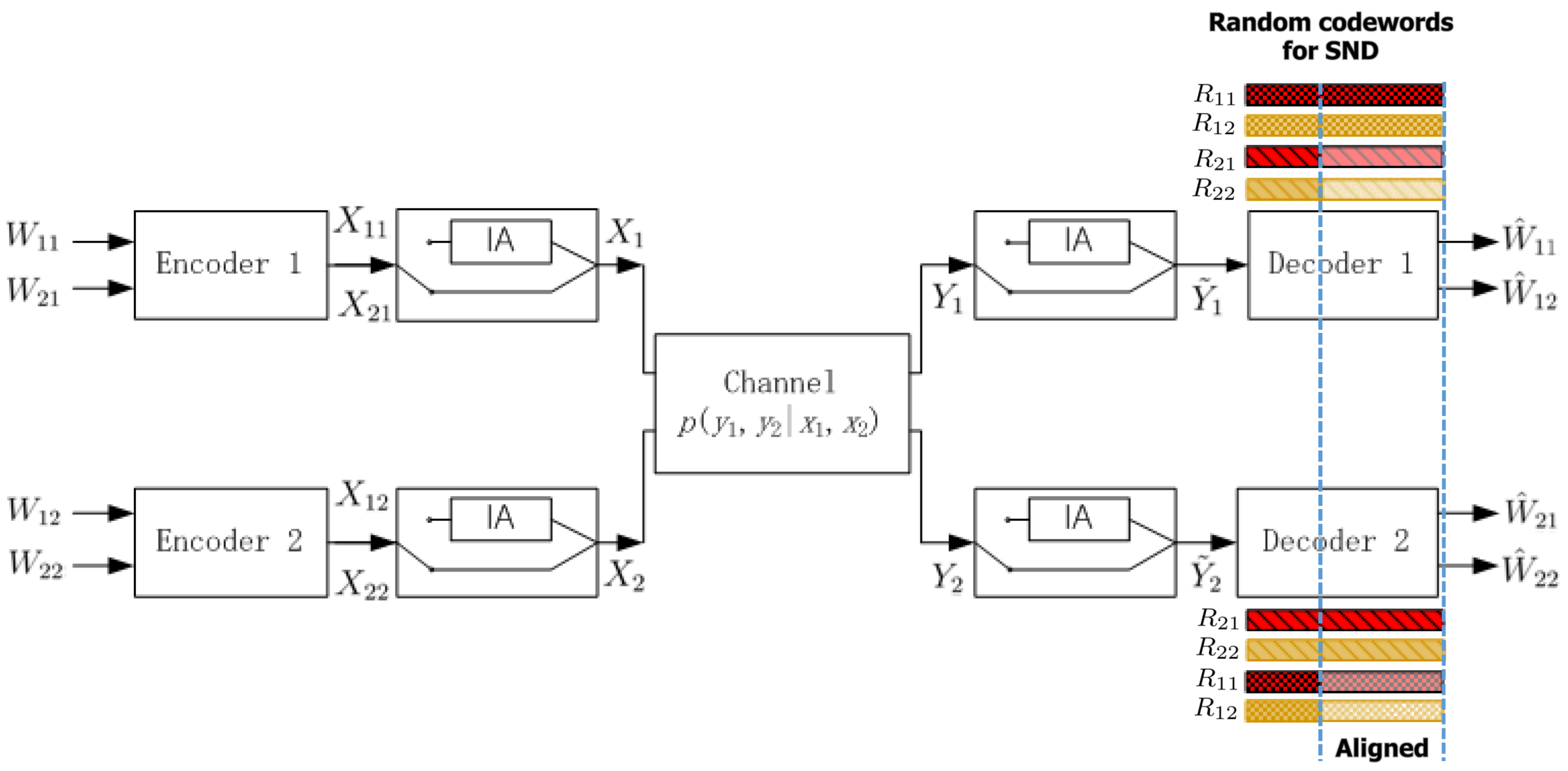}\\
  \caption{Simultaneous non-unique decoding the received signals in 2-user X-channel}
  \label{fig:onoff5}
\end{figure}
\item[(3)]Transmitting the beamformed signals: The beamformed signals at each transmitter are transmitted. 
Note that the relatively dark-shaded parts of each codeword, in Fig. \ref{fig:onoff3}, indicate the portion of the precoded. 
\item[(4)]On-off switched beamforming at the receivers:
When the receivers receive codewords,  the codewords are postprocessed by switching IA beamformer on/off according to the predetermined on-off pattern  as shown in Fig. \ref{fig:onoff4}.  

\item[(5)]Simultaneous non-unique decoding:
Finally,  simultaneous non-unique decoding is performed after the whole codeword is received at each receiver. As illustrated in Fig. \ref{fig:onoff5}, the parts marked as `Aligned' in the codewords represent the beamformed portion during which interfering signals are aligned.
 \end{enumerate}
Based on the described process of the on-off switched IA above, we justify the intermittent utilization of IA for the DMT improvement in the following subsections.

\subsection{DMT of On-off Switched Interference Alignment with Symbol Extension}

The IA scheme for on-off switching requires  symbol extension in order to align the interfering
signals in the same dimensional space. For the 2-user MIMO X
channel with two antennas each, each transmitter sends two symbols to each receiver over three symbol times. Consequently, each user can achieve $\frac{4}{3}$ DoF by decoding 4
		symbols over three symbol times. With the time extension, the received signal at receiver $i$ is rewritten as
$\bar{\mathbf{y}}^{[i]}=\sum_{j=1,2}\bar{\mathbf{H}}^{[ij]}\bar{\mathbf{x}}^{[j]}+\bar{\mathbf{z}}^{[i]}$
where $\bar{\mathbf{y}}^{[i]}=[\bar{y}_1^{[i]}~~\bar{y}_2^{[i]}~~\cdots~~\bar{y}_6^{[i]}]^T\in \mathcal{C}^{6\times 1}$,
$\bar{\mathbf{x}}^{[j]}=[\bar{x}_1^{[j]}~~\bar{x}_2^{[j]}~~\cdots~~\bar{x}_6^{[j]}]^T
\in \mathcal{C}^{6\times 1}$, $\bar{\mathbf{z}}^{[i]}= [\bar{z}_1^{[i]}~~\cdots~~\bar{z}_6^{[i]}]^T$, and
\begin{align}
\bar{\mathbf{H}}^{[ij]} = \left[
                  \begin{array}{cccccc}
                    h_{11}^{[ij]} & h_{12}^{[ij]} &0&0&0&0\\
                    h_{21}^{[ij]}  & h_{22}^{[ij]}&0&0&0&0 \\
                    0&0&h_{11}^{[ij]} & h_{12}^{[ij]} &0&0\\
                    0&0&h_{21}^{[ij]}  & h_{22}^{[ij]}&0&0 \\
                    0&0&0&0&h_{11}^{[ij]} & h_{12}^{[ij]} \\
                    0&0&0&0&h_{21}^{[ij]}  & h_{22}^{[ij]} \\
                  \end{array}
                \right].
\end{align}
With local CSIT only, we design a two-stage beamforming scheme to align the interfering signals. The role of the beamforming matrices at the first stage is to easily cancel out the interfering signals aligned by the second stage beamforming matrices. Then, with a power constraint $P$, the transmit signal of transmitter $j$ is constituted by
\begin{align}
\bar{\mathbf{x}}^{[j]} = \mathbf{V}^{[1j]}\mathbf{U}^{[1]}\mathbf{s}^{[1j]}+\mathbf{V}^{[2j]}\mathbf{U}^{[2]}\mathbf{s}^{[2j]}
\end{align}  where the transmit signal from transmitter $j$ to receiver $i$ is given by $\mathbf{s}^{[ij]}=[s_1^{[ij]}~~s_2^{[ij]}]^T$, $\mathbf{U}^{[i]}$ denotes the first stage beamforming matrix for receiver $i$, and $\mathbf{V}^{[ij]}$ is the second stage beamforming matrix for receiver $i$. In our interference alignment, the first stage beamforming matrices are designed as
\begin{align}
\mathbf{U}^{[1]} = \left[
                  \begin{array}{cc}
                    1&0\\
                    1&0 \\
                    0&0\\
                    0&1 \\
                    0&1 \\
                    0&0\\
                  \end{array}
                \right],~~~~~\mathbf{U}^{[2]} = \left[
                  \begin{array}{cc}
                    1&0\\
                    0&0 \\
                    1&0\\
                    0&1 \\
                    0&0 \\
                    0&1\\
                  \end{array}
                \right].
\end{align} To align interfering signals in the same dimensional space, the following conditions have to be satisfied:
\begin{align}
\bar{\mathbf{H}}^{[11]}\mathbf{V}^{[21]}\mathbf{U}^{[2]} = \bar{\mathbf{H}}^{[12]}\mathbf{V}^{[22]}\mathbf{U}^{[2]},\\
\bar{\mathbf{H}}^{[21]}\mathbf{V}^{[11]}\mathbf{U}^{[1]} = \bar{\mathbf{H}}^{[22]}\mathbf{V}^{[12]}\mathbf{U}^{[1]}.
\end{align} Given the conditions of interference alignment, the second stage beamforming matrices can be obtained as
\begin{align}
\mathbf{V}^{[ij]} = c^{[ij]}(\bar{\mathbf{H}}^{[kj]})^{-1}, ~~~~~~i\neq k
\end{align}
where $c^{[ij]}=1/\|(\bar{\mathbf{H}}^{[kj]})^{-1}\|$. Then, the received signal at receiver $1$ becomes
\begin{align}
\bar{\mathbf{y}}^{[1]}&=\bar{\mathbf{H}}^{[11]}\mathbf{V}^{[11]}\mathbf{U}^{[1]}\mathbf{s}^{[11]}+\bar{\mathbf{H}}^{[11]}\mathbf{V}^{[21]}\mathbf{U}^{[2]}\mathbf{s}^{[21]}\nonumber\\
&+\bar{\mathbf{H}}^{[12]}\mathbf{V}^{[12]}\mathbf{U}^{[1]}\mathbf{s}^{[12]}+\bar{\mathbf{H}}^{[12]}\mathbf{V}^{[22]}\mathbf{U}^{[2]}\mathbf{s}^{[22]}+\bar{\mathbf{z}}^{[1]}\nonumber\\
&=\!c^{[11]}\!\bar{\mathbf{H}}^{[11]}\!(\bar{\mathbf{H}}^{[21]}\!)\!^{-1}\!\mathbf{U}^{[1]}\!\mathbf{s}^{[11]}\!\!+\!c^{[12]}\!\bar{\mathbf{H}}^{[12]}\!(\bar{\mathbf{H}}^{[22]}\!)\!^{-1}\!\mathbf{U}^{[1]}\!\mathbf{s}^{[12]}\nonumber\\
&+\mathbf{U}^{[2]}(c^{[21]}\mathbf{s}^{[21]}+c^{[22]}\mathbf{s}^{[22]})+\bar{\mathbf{z}}^{[1]}\nonumber
\end{align} and  the corresponding $6\times6$  effective channel for receiver 1 can be rewritten as
\begin{align}
[\bar{\mathbf{H}}^{[11]}(\bar{\mathbf{H}}^{[21]})^{-1}\mathbf{U}^{[1]}~~\bar{\mathbf{H}}^{[12]}(\bar{\mathbf{H}}^{[22]})^{-1}\mathbf{U}^{[1]}~~\mathbf{U}^{[2]}].
\end{align}

 To cancel out the interfering signals aligned in the same dimensional space, we subtract $\bar{y}_3^{[1]}$ and $\bar{y}_6^{[1]}$ from $\bar{y}_1^{[1]}$ and $\bar{y}_4^{[1]}$, respectively, as follows
\begin{align}
\left[\begin{array}{c}
                    \bar{y}_1^{[1]}-\bar{y}_3^{[1]}\\
                    \bar{y}_2^{[1]} \\
                    \bar{y}_4^{[1]}-\bar{y}_6^{[1]}\\
                    \bar{y}_5^{[1]} \\
                  \end{array}
                \right] = \tilde{\mathbf{H}}\left[\begin{array}{c}
                    \mathbf{s}^{[11]}\\
                    \mathbf{s}^{[12]}\\
                  \end{array}
                \right]+\left[\begin{array}{c}
                    \bar{z}_1^{[1]}-\bar{z}_3^{[1]}\\
                    \bar{z}_2^{[1]} \\
                    \bar{z}_4^{[1]}-\bar{z}_6^{[1]}\\
                    \bar{z}_5^{[1]} \\
                  \end{array}\right]
\end{align}
where $\tilde{\mathbf{H}}$ represents the effective channel matrix with full rank of $4$ after the interfering signals are removed. With a zero-forcing (ZF) decoder, the achievable rate is obtained as
\begin{align}
R_1 = R_{11}+R_{12}&\le\frac{1}{3}\log\det(I+\tilde{\rho}\tilde{\mathbf{H}}\tilde{\mathbf{H}}^{\dagger}),
\end{align}
where $\tilde{\rho}$ denotes effective SNR changed by the conventional IA.
Similarly, the rate constraint for $R_2 = R_{21}+R_{22}$ can be readily obtained.

\begin{theorem}
  When on-off switching is applied to the IA with symbol extension and simultaneous non-unique decoding is performed at each receiver, 
  DMT of the 2-user two-antenna X-channel with local CSIT is
  obtained as
\begin{align}
\label{Eq:DMTTS}
d_{IA}(r)&=\min\Bigg\{
\min\Big(\frac{2}{a+3},\frac{1}{4a}\Big)(6-3r-2a),\nonumber\\&\quad~~~~~~~\min\Big(\frac{4}{a+3},\frac{1}{4a}\Big)(6-6r+2a),d^*_{2,2}(r)\Bigg\}
\end{align}
where $a \in [0,1]$ is a given portion for IA usage.
\end{theorem}

\begin{IEEEproof}
We formulate the achievable rate region to represent outage events.
In view of the received signal at each receiver, the existence of interference depends on whether the IA is switched on or off.
That is, the portion $a$ of a received codeword is free from  interference. Consequently, the achievable rate regions obtained
by simultaneous non-unique decoding are different whether the IA is switched on or off.
Note that we consider a quasi-static channel as assumed in Section II, which means the codeword experiences an unchanged channel state (or gain) once after the channel has been realized, regardless that the IA is switched on or off.  As such,  for a given channel state $S=s$,  the maximum of achievable rate is rigorously represented as 
$C = \max_{p(x|s)} I(X;Y | S=s)$, where $S$ is the channel state and $s$ is one of possible channel states.  However,
since for a given time invariant channel gain the channel can be interpreted as a normalized AWGN channel, 
with a slight abuse of notation, we analyze the achievable rate region  based on  $I(X;Y)$. 

Let $\mathcal{R}_i^{\textrm{wIA}}$ and $\mathcal{R}_i^{\textrm{woIA}}$ be achievable rate region by simultaneous non-unique decoding of user $i$ with and without
IA based beamformer, respectively. 
As explained in Section III. B, not globally perfect CSIT but local CSIT is required for IA based beamformer.
In the case when the IA based beamformer is switched on, $\mathcal{R}_{i,\textrm{IAN}}^{\textrm{wIA}}\supset\mathcal{R}_{i,\textrm{SD}}^{\textrm{wIA}}$, $\forall i=1,2$ since
the rate region of simultaneous decoding is bottlenecked by the sum rate constraint, although interference is nulled out by IA.
Accordingly,
\begin{align}
\mathcal{R}^{\textrm{wIA}}=\mathcal{R}_1^{\textrm{wIA}}\cap\mathcal{R}_2^{\textrm{wIA}}=\mathcal{R}_{1,\textrm{IAN}}^{\textrm{wIA}}\cap\mathcal{R}_{2,\textrm{IAN}}^{\textrm{wIA}}
\end{align}
which represents the set of rate pairs $(R_1^{\textrm{wIA}},R_2^{\textrm{wIA}})$ satisfying
\begin{align}
R_i^{\textrm{wIA}}&\leq I(X_{i1},X_{i2};\tilde{Y_{i}}),\forall i=1,2
\end{align}
where $\tilde{Y_{i}}$, $\forall i=1,2$ denotes the beamformed output of user $i$ when the beamformer blocks of each user are switched  on.

In the case when IA is switched off,
\begin{align}
\mathcal{R}^{\textrm{woIA}}&=\mathcal{R}_1^{\textrm{woIA}}\cap\mathcal{R}_2^{\textrm{woIA}}\nonumber\\
&=(\mathcal{R}_{1,\textrm{IAN}}^{\textrm{woIA}}\cup\mathcal{R}_{1,\textrm{SD}}^{\textrm{woIA}})\cap(\mathcal{R}_{2,\textrm{IAN}}^{\textrm{woIA}}\cup\mathcal{R}_{2,\textrm{SD}}^{\textrm{woIA}})\\
&=(\mathcal{R}_{1,\textrm{IAN}}^{\textrm{woIA}}\cap\mathcal{R}_{2,\textrm{IAN}}^{\textrm{woIA}})\cup
(\mathcal{R}_{1,\textrm{IAN}}^{\textrm{woIA}}\cap\mathcal{R}_{2,\textrm{SD}}^{\textrm{woIA}})\nonumber\\
&~~\cup
(\mathcal{R}_{1,\textrm{SD}}^{\textrm{woIA}}\cap\mathcal{R}_{2,\textrm{IAN}}^{\textrm{woIA}})\cup
(\mathcal{R}_{1,\textrm{SD}}^{\textrm{woIA}}\cap\mathcal{R}_{2,\textrm{SD}}^{\textrm{woIA}})
\end{align} which represents
the set of rate pairs $(R_1^{\textrm{woIA}},R_2^{\textrm{woIA}})$ satisfying
\begin{align}
R_i^{\textrm{woIA}}&\leq I(X_{i1},X_{i2};Y_{i}),\forall i=1,2 ~~~\textrm{or}
\end{align}
\begin{align}
R_1^{\textrm{woIA}}&\leq I(X_{11},X_{12};Y_{1}),\\
R_i^{\textrm{woIA}}&\leq I(X_{i1},X_{i2};Y_{2}|X_{\bar{i}1},X_{\bar{i}2}),\forall i=1,2, i\neq\bar{i}\\
R_1^{\textrm{woIA}}+R_2^{\textrm{woIA}}&\leq I(X_{11},X_{12},X_{21},X_{22};Y_{2}), ~~~\textrm{or}
\end{align}
\begin{align}
R_i^{\textrm{woIA}}&\leq I(X_{i1},X_{i2};Y_{1}|X_{\bar{i}1},X_{\bar{i}2}),\forall i=1,2, i\neq\bar{i}\\
R_1^{\textrm{woIA}}+R_2^{\textrm{woIA}}&\leq I(X_{11},X_{12},X_{21},X_{22};Y_{1}),\\
R_2^{\textrm{woIA}}&\leq I(X_{21},X_{22};Y_{2}),~~~\textrm{or}
\end{align}
\begin{align}
R_1^{\textrm{woIA}}&\leq I(X_{11},X_{12};Y_{j}|X_{21},X_{22}),\\
R_2^{\textrm{woIA}}&\leq I(X_{21},X_{22};Y_{j}|X_{11},X_{12}),\\
R_1^{\textrm{woIA}}+R_2^{\textrm{woIA}}&\leq I(X_{11},X_{12},X_{21},X_{22};Y_{j}),\forall j =1,2.
\end{align}

Since $a R_i^{\textrm{wIA}}+(1-a)R_i^{\textrm{woIA}}=R_i$, $ i=1,2$
and a symmetric target rate for each user, i.e., $R_1=R_2=R$, is assumed, 
the outage event of the on-off switched IA scheme with simultaneous non-unique decoding is obtained by
\begin{align}
&\mathcal{E}_{IA}=\big\{ aI(X_{i1},X_{i2};\tilde{Y_{i}})+(1-a)I(X_{i1},X_{i2};Y_{i})<R,\nonumber\\
&~~~~\exists i=1,2\big\}~~\nonumber\\
\cap&\big\{\big\{aI(X_{11},X_{12};\tilde{Y_{1}})+(1-a)I(X_{11},X_{12};Y_{1})<R\big\} ~~ \nonumber\\
&\cup \big\{aI(X_{i1},X_{i2};\tilde{Y_{i}})\nonumber\\
&~+(1-a)I(X_{i1},X_{i2};Y_{2}|X_{\bar{i}1},X_{\bar{i}2})<R,\exists i=1,2, i\neq\bar{i}\big\} ~~ \nonumber\\
&\cup \big\{a\sum_{i=1,2}I(X_{i1},X_{i2};\tilde{Y_{i}})\nonumber\\
&~+(1-a)I(X_{11},X_{12},X_{21},X_{22};Y_{2})<2R\big\}\big\} ~~\nonumber\\
\cap&\big\{\big\{aI(X_{i1},X_{i2};\tilde{Y_{i}})\nonumber\\
&~+(1-a)I(X_{i1},X_{i2};Y_{1}|X_{\bar{i}1},X_{\bar{i}2})<R,\exists i=1,2, i\neq\bar{i} \big\}~~ \nonumber\\
&\cup \big\{a\sum_{i=1,2}I(X_{i1},X_{i2};\tilde{Y_{i}})\nonumber\\
&~+(1-a)I(X_{11}\!,X_{12},X_{21},X_{22};Y_{1}) <2R\big\} ~~ \nonumber\\
&\cup \big\{aI(X_{21},X_{22};\tilde{Y_{2}})+(1-a)I(X_{21},X_{22};Y_{2}) <R \big\}\big\}~~ \nonumber\\
\cap&\big\{\big\{aI(X_{11}\!,X_{12};\!\tilde{Y_{1}})\!+\!(1\!-a)I(X_{11}\!,X_{12};Y_{j}|X_{21}\!,X_{22})\!<\!R\big\} ~~ \nonumber\\
&\cup\! \big\{aI(X_{21}\!,\!X_{22};\tilde{Y_{2}})\!+\!(1\!-a)I(X_{21}\!,\!X_{22};Y_{j}|X_{11}\!,\!X_{12})\!<\!R\big\} ~~ \nonumber\\
&\cup \big\{a\sum_{i=1,2}I(X_{i1},X_{i2};\tilde{Y_{i}})\nonumber\\
&~+(1-a)I(X_{11},X_{12},X_{21},X_{22};Y_{j}) <2R,~\exists j =1,2\big\}\big\}.
\end{align}

Take note that the terms, $I(X_{11},X_{12};Y_{1})$ and $I(X_{21},X_{22};Y_{2})$ are constant when SNR is sufficiently high
because interference grows with SNR and is dealt as noise. Consequently,
\begin{align}
\label{Eq:IAout}
&\textrm{Pr}\{\mathcal{E}_{IA}\}\doteq \textrm{Pr}\{\mathcal{E}_1\cap\mathcal{E}_2\cap\mathcal{E}_3\cap\mathcal{E}_4\},~~\textrm{where}\nonumber\\
&\mathcal{E}_1=\big\{\big\{aI(X_{11},X_{12};\tilde{Y_{1}})\!<\!R\big\}\!\cup\!\big\{ aI(X_{21},X_{22};\tilde{Y_{2}})\!<\!R\big\}\big\},~~\nonumber\\
&\mathcal{E}_2=\big\{\big\{aI(X_{i1},X_{i2};\tilde{Y_{i}})\nonumber\\
&~~~~+(1-a)I(X_{i1},X_{i2};Y_{2}|X_{\bar{i}1},X_{\bar{i}2})\!<\!R,\exists i=1,2, i\neq\bar{i}\big\} ~~ \nonumber\\
&~~~\cup \big\{a\sum_{i=1,2}I(X_{i1},X_{i2};\tilde{Y_{i}})\nonumber\\
&~~~~+(1-a)I(X_{11},X_{12},X_{21},X_{22};Y_{2})\!<\!2R\big\}\big\}, ~~\nonumber\\
&\mathcal{E}_3=\big\{\big\{aI(X_{i1},X_{i2};\tilde{Y_{i}})\nonumber\\
&~~~~+(1-a)I(X_{i1},X_{i2};Y_{1}|X_{\bar{i}1},X_{\bar{i}2})\!<\!R,\exists i=1,2, i\neq\bar{i}\big\} ~~ \nonumber\\
&~~~\cup \big\{a\sum_{i=1,2}I(X_{i1},X_{i2};\tilde{Y_{i}})\nonumber\\
&~~~~+(1-a)I(X_{11},X_{12},X_{21},X_{22};Y_{1}) \!<\!2R\big\} \big\},~~ \nonumber\\
&\mathcal{E}_4\!\!=\!\big\{\!\big\{\!aI(X_{11}\!,\!X_{12};\tilde{Y_{1}})\!+\!(1\!-\!a)I(X_{11}\!,\!X_{12};Y_{j}|X_{21}\!,\!X_{22})\!\!<\!\!R \big\} \nonumber\\
&~~\cup\! \big\{aI(X_{21}\!,\!X_{22};\tilde{Y_{2}})\!+\!(1\!-\!a)I(X_{21}\!,\!X_{22};Y_{j}|X_{11}\!,\!X_{12})\!\!<\!\!R\big\} ~~ \nonumber\\
&~~\cup \!\big\{a\sum_{i=1,2}I(X_{i1},X_{i2};\tilde{Y_{i}})\nonumber\\
&~~~~+(1-a)I(X_{11},X_{12},X_{21},X_{22};Y_{j}) \!<\!2R,~\exists j =1,2\big\}\big\}.\nonumber
\end{align} 
Since $\mathcal{E}_1$, $\mathcal{E}_2$, and $\mathcal{E}_3$ are definitely included in $\mathcal{E}_4$, the dominant scale of $\textrm{Pr}\{\mathcal{E}_{IA}\}$
is bottlenecked by $\textrm{Pr}\{\mathcal{E}_4\}$ as
\begin{align}
\textrm{Pr}\{\mathcal{E}_{IA}\}&\doteq\textrm{Pr}\{\mathcal{E}_4 \}\nonumber\\
&=\textrm{Pr}\big\{\big\{aI(X_{j1},X_{j2};\tilde{Y_{j}})\nonumber\\&~~~~+(1-a)I(X_{j1},X_{j2};Y_{j}|X_{k1},X_{k2})<R\big\} ~~ \nonumber\\
&~\cup \big\{a\sum_{i=1,2}I(X_{i1},X_{i2};\tilde{Y_{i}})\nonumber\\&~~~~+(1-a)I(X_{11},X_{12},X_{21},X_{22};Y_{j}) <2R,\nonumber\\&~~~~~\exists j,k =1,2, j\neq k \big\}\big\}\nonumber\\
& \overset{(a)}{\doteq}\textrm{Pr}\big\{\big\{\frac{4}{3}a\log\left(1+\rho\left|h_{jk}^{[ii]}\right|^2\right)\nonumber\\&~~~~~~+(1-a)\log\det\left(\mathbf{I}+\rho\mathbf{H}^{[ii]}\mathbf{H}^{[ii]\dagger}\right)<R\big\} ~~ \nonumber\\
&~\cup \big\{\sum_{l=1}^{2} \frac{4}{3}a\log\left(1+\rho\left|h_{jk}^{[ll]}\right|^2\right)\nonumber\\&~~~~~~+(1-a)\log\det\left(\mathbf{I}+\rho\bar{\mathbf{H}}\bar{\mathbf{H}}^{\dagger}\right) <2R,\nonumber\\&~~~~~~~\exists i,j,k\in\{1,2\} \big\}\big\},\nonumber
\end{align}
where $\bar{\mathbf{H}}=\left[\mathbf{H}^{[i1]}\mathbf{H}^{[i2]}\right]$ and (a) is because each user achieves diversity gain  of 1 and DoF of $\frac{4}{3}$ with the IA in the 2-user X channel with two antennas \cite{IAAla}.
Because all  the channels of the two symmetric users are \emph{i.i.d.}, indices $i, j,$  and $k$ do not affect the dominant scale of $\textrm{Pr}\{\mathcal{E}_{IA}\}$.
For that reason, let us consider the $i=j=k=1$ case for simple notation.

Let
$\mathcal{E}_{IA_1}=\Big\{\frac{4}{3}a\log$$\left(1+\rho\left|h_{11}^{[11]}\right|^2\right)\\+(1-a)\log\det\left(\mathbf{I}+\rho\mathbf{H}^{[11]}\mathbf{H}^{[11]\dagger}\right)<R \Big\}, \\
\mathcal{E}_{IA_2}
=\Big\{\sum_{l=1,2}\frac{4}{3}a\log\left(1+\rho\left|h_{11}^{[ll]}\right|^2\right)\\+(1-a)\log\det\left(\mathbf{I}+\rho\bar{\mathbf{H}}\bar{\mathbf{H}}^{\dagger}\right)
<2R \Big\}$,
and $v_{nm}^{[ij]}$ be the exponential order
of $1/\left|h_{nm}^{[ij]}\right|^2$, $\forall i,j,m,n=1,2$. Then, the probability for $\mathcal{E}_{IA_1}$ is represented as
\begin{align}
\textrm{Pr}\{\mathcal{E}_{IA_1}\}\!\!&=\!\textrm{Pr}\Big\{\frac{4}{3}a\log\left(1+\rho\left|h_{11}^{[11]}\right|^2\right)\nonumber\\&\quad+\!(1-a)\log\det\!\left(\mathbf{I}+\rho\mathbf{H}^{[11]}\mathbf{H}^{[11]\dagger}\right)
<r\log\rho \Big\}\nonumber\\
&\doteq\!\textrm{Pr}\Big\{\frac{4}{3}a\log\rho\!\!\left|h_{11}^{[11]}\right|^2\!\!\!\!+\!\!(1-a)\sum_{i=1}^{2}\log\rho\lambda_i\!\!<\!r\log\rho \Big\}\nonumber\\
&=\!\textrm{Pr}\Big\{\frac{4}{3}a\log\rho\!\left|h_{11}^{[11]}\right|^2\!\!\!\!+\!\!(1-a)\log\rho^2\lambda_1\!\lambda_2\!\!<\!r\log\rho \Big\}\nonumber\\
&=\textrm{Pr}\Big\{\frac{4}{3}a\log\rho\left|h_{11}^{[11]}\right|^2\!\!\!\!+\!(1-a)\log\rho^2\nonumber\\&\quad\times\!(\mathbf{G}_{(1,1)}^{[11]}\mathbf{G}_{(2,2)}^{[11]}\!\!-\!\!\mathbf{G}_{(1,2)}^{[11]}\mathbf{G}_{(2,1)}^{[11]})<r\log\rho \Big\},
\end{align}
where $\lambda_i$ for $i=1,2$ are eigenvalues of $\mathbf{H}^{[11]}\mathbf{H}^{[11]\dagger}$, $\mathbf{G}^{[11]}=\mathbf{H}^{[11]}\mathbf{H}^{[11]\dagger}$, and
$\mathbf{G}_{(n,m)}^{[11]}$ denotes the n-th row and m-th column element of $\mathbf{G}^{[11]}$. Consequently,
\begin{align}
\label{Eq:ets1}
\textrm{Pr}\{\mathcal{E}_{IA_1}\}&\doteq\textrm{Pr}\Big\{\frac{4}{3}a\log\rho\left|h_{11}^{[11]}\right|^2+(1-a)\log\rho^2\nonumber\\&\quad\times\!\left(\left|h_{11}^{[11]}\right|^2\!\left|h_{22}^{[11]}\right|^2\!\!\!+\left|h_{12}^{[11]}\right|^2\!\!\left|h_{21}^{[11]}\right|^2\right)\!<\!r\log\rho \Big\}\nonumber\\
&\doteq\textrm{Pr}\Big\{\frac{4}{3}a\log\rho^{1-v_{11}^{[11]}}+(1-a)\nonumber\\&\quad\times\log(\rho^{2-v_{11}^{[11]}-v_{22}^{[11]}}+\rho^{2-v_{12}^{[11]}-v_{21}^{[11]}})<r\log\rho \Big\}\nonumber\\
&=\textrm{Pr}\Big\{\log\rho^{\frac{4a}{3(1-a)}(1-v_{11}^{[11]})}\nonumber\\&\quad\times(\rho^{2-v_{11}^{[11]}-v_{22}^{[11]}}\!+\!\rho^{2-v_{12}^{[11]}-v_{21}^{[11]}})\!<\!\log\rho^{\frac{r}{1-a}} \Big\}\nonumber\\
&=\textrm{Pr}\Big\{\rho^{\frac{4a}{3(1-a)}(1-v_{11}^{[11]})+2-v_{11}^{[11]}-v_{22}^{[11]}}\nonumber\\&\quad+\rho^{\frac{4a}{3(1-a)}(1-v_{11}^{[11]})+2-v_{12}^{[11]}-v_{21}^{[11]}}<\rho^{\frac{r}{1-a}} \Big\}\nonumber\\
&\doteq\textrm{Pr}\Bigg\{\max\Big\{\frac{4a(1-v_{11}^{[11]})}{3(1-a)}\!+\!2\!-v_{11}^{[11]}\!-v_{22}^{[11]},\nonumber\\
&\qquad\qquad\frac{4a(1-v_{11}^{[11]})}{3(1-a)}\!+\!\!2\!-\!v_{12}^{[11]}\!-\!v_{21}^{[11]}\Big\}\!<\!\frac{r}{1-a}\!\Bigg\}
\end{align}


The probability for $\mathcal{E}_{IA_2}$ is given by
\begin{align}
\textrm{Pr}\{\mathcal{E}_{IA_2}\}&=\textrm{Pr}\Big\{\sum_{l=1,2}\frac{4}{3}a\log\left(1+\rho\left|h_{11}^{[ll]}\right|^2\right)+(1-a)\nonumber\\&\quad\times\log\det\left(1+\rho\bar{\mathbf{H}}\bar{\mathbf{H}}^{\dagger}\right)
<2r\log\rho \Big\}\nonumber\\
&\doteq\textrm{Pr}\Big\{\!\!\sum_{l=1,2}\frac{4}{3}a\log\rho\left|h_{11}^{[ll]}\right|^2\!\!\!\nonumber\\&\quad+\!(1-a)\log\rho^2\bar{\lambda_1}\bar{\lambda_2}<2r\log\rho \Big\}\nonumber\\
&=\textrm{Pr}\Big\{\sum_{l=1,2}\frac{4}{3}a\log\rho\left|h_{11}^{[ll]}\right|^2+(1-a)\log\rho^2\nonumber\\&\quad\times(\bar{\mathbf{G}}_{(1,1)}\bar{\mathbf{G}}_{(2,2)}\!\!-\!\!\bar{\mathbf{G}}_{(1,2)}\bar{\mathbf{G}}_{(2,1)}\!)\!<\!2r\log\rho \!\Big\}\!\nonumber
\end{align}
where $\bar{\lambda_i}$ for $i=1,2$ are eigenvalues of $\bar{\mathbf{H}}\bar{\mathbf{H}}^{\dagger}$, $\bar{\mathbf{G}}=\bar{\mathbf{H}}\bar{\mathbf{H}}^{\dagger}$, and
$\bar{\mathbf{G}}_{(n,m)}$ represents the n-th row and m-th column element of $\bar{\mathbf{G}}$. Calculating $\bar{\mathbf{G}}_{(1,1)}\bar{\mathbf{G}}_{(2,2)}-\bar{\mathbf{G}}_{(1,2)}\bar{\mathbf{G}}_{(2,1)}$,
the dominant scale of the outage probability of $\mathcal{E}_{IA_2}$ is obtained as
(\ref{Eq:ets2}) on the next page.
\begin{figure*}
\begin{align}
\label{Eq:ets2}
\textrm{Pr}\{\mathcal{E}_{IA_2}\}&\doteq\textrm{Pr}\Big\{\frac{4a}{3(1-a)}\log\rho^{2-v_{11}^{[11]}-v_{11}^{[22]}}+\log\!\!\!\sum_{l_1,l_2,k_1,k_2\in \{1,2\}\bigcap(l_1,k_1)\neq(l_2,k_2)}\rho^{2-v_{1k_1}^{[1l_1]}-v_{2k_2}^{[1l_2]}}<2r\log\rho \Big\}\nonumber\\
&\doteq\textrm{Pr}\Big\{\sum_{l_1,l_2,k_1,k_2\in \{1,2\}\bigcap(l_1,k_1)\neq(l_2,k_2)}\rho^{\frac{4a}{3(1-a)}(2-v_{11}^{[11]}-v_{11}^{[22]})+2-v_{1k_1}^{[1l_1]}-v_{2k_2}^{[1l_2]}}<\rho^{\frac{2r}{1-a}} \Big\}\nonumber\\
&\doteq\textrm{Pr}\Big\{\max_{l_1,l_2,k_1,k_2\in \{1,2\}\bigcap(l_1,k_1)\neq(l_2,k_2)}\rho^{\frac{4a}{3(1-a)}(2-v_{11}^{[11]}-v_{11}^{[22]})+2-v_{1k_1}^{[1l_1]}-v_{2k_2}^{[1l_2]}}<\rho^{\frac{2r}{1-a}} \Big\}.
\end{align}
\hrulefill
\end{figure*}
The dominant scale of $\textrm{Pr}\{\mathcal{E}_{IA_1}\}$ and $\textrm{Pr}\{\mathcal{E}_{IA_2}\}$ can be found by considering all the cases in (\ref{Eq:ets1}) and (\ref{Eq:ets2}), respectively.
See Appendix A for the proof of finding the dominant scale of $\textrm{Pr}\{\mathcal{E}_{IA_1}\}$ and $\textrm{Pr}\{\mathcal{E}_{IA_2}\}$.
\end{IEEEproof}

  \begin{remark}
  The DMT improvement via on-off switched IA, in \emph{Theorem 1}, mainly comes from the fact that the codewords undergo both the interference aligned and  non-aligned conditions before decoding, which retains the features of both IA and SD, owing to the simultaneous non-unique decoder. 
As a result, the on-off switched IA beamformer with simultaneous non-unique decoder can achieve the rate region expanded from the \emph{union} of the rate regions of IAN and SD via the coded time-shared methodology.
 Therefore, the DMT performance of the proposed technique is not bottlenecked by either IAN or SD but is substantially improved in terms of DMT. 
  Leveraging the merit by optimizing the portion $a$ of IA utilization, diversity gain of 4 as well as multiplexing gain per user of $\frac{4}{3}$ can be achieved. Details on optimization of $a$ are dealt in Section IV.
  \end{remark}



%

\subsection{DMT of On-off Switched Interference Alignment with Alamouti Coding}
\begin{theorem} When on-off switching is applied to the IA with Alamouti coding scheme and simultaneous non-unique decoding is performed at each receiver, 
  DMT of the 2-user two-antenna X-channel under local CSIT is derived as
\begin{align}
\label{Eq:DMTTSA}
d_{IAA}(r)&=\min\!\!\Bigg\{\!
\min\Big(\!
\frac{2}{a+3},\frac{1}{2a}\Big)(6-3r-2a),\nonumber\\&\quad~~~~~~~\min\!\Big(\frac{4}{a+3},\frac{1}{2a}\Big)(6-6r+2a),d^*_{2,2}(r)\Bigg\}
\end{align}
where $a ~\in [0,1]$ is a given time portion for the IA with Alamouti coding.
\end{theorem}

\begin{IEEEproof}
We prove achievable DMT of the 2-user 2-antenna X-channel when the on-off switched IA with Alamouti coding is adopted and simultaneous non-unique decoding is performed at each receiver. The details of IA with Alamouti coding are referred to \cite{IAAla}.

Similar to the case of the proposed scheme based on the IA with symbol extension,
the dominant scale of an outage event for the on-off switched IA with Alamouti coding scheme when simultaneous non-unique decoding is performed at each receiver is given by
\begin{align}
\textrm{Pr}\{\mathcal{E}_{IAA}\}&\doteq \textrm{Pr}\big\{\big\{aI(X_{j1},X_{j2};\tilde{Y_{j}})\nonumber\\
&~~~+(1-a)I(X_{j1},X_{j2};Y_{j}|X_{k1},X_{k2})<R \big\}~~ \nonumber\\
\cup \big\{ a\!\!\sum_{i=1,2}&I(X_{i1}\!,\!X_{i2};\tilde{Y_{i}})\!+\!(1\!-\!a)I(X_{11}\!,\!X_{12}\!,\!X_{21}\!,\!X_{22};Y_{j}) \nonumber\\
&<2R,~\forall j,k =1,2, j\neq k \big\}\big\}\nonumber\\
&\doteq\textrm{Pr}\left\{\mathcal{E}_{IAA_1}\cup\mathcal{E}_{IAA_2}\right\},~~\textrm{where}\nonumber\\
\mathcal{E}_{IAA_1}=\bigg\{&\frac{4}{3}a\log\left(1+\rho\left(\left|h_{jk}^{[i1]}\right|^2+\left|h_{jk}^{[i2]}\right|^2\right)\right)\nonumber\\
&~~~+(1-a)\log\det\left(\mathbf{I}+\rho\mathbf{H}^{[ii]}\mathbf{H}^{[ii]\dagger}\right)<R \bigg\}, \nonumber\\
\mathcal{E}_{IAA_2}\!\!=\!\!\bigg\{\!&\sum_{l=1,2}\! \frac{4}{3}a\log\!\left(\!1\!\!+\!\!\rho\!\left(\left|h_{jk}^{[l1]}\right|^2\!\!\!+\!\!\left|h_{jk}^{[l2]}\right|^2\right)\right)\!\!\nonumber\\
&+\!\!(\!1\!-\!a\!)\!\log\!\det\left(\mathbf{I}+\rho\bar{\mathbf{H}}\bar{\mathbf{H}}^{\dagger}\right)\!\! <\!2R,\forall i\!,j\!,k\!\!\in\!\{1,2\} \!\bigg\},\nonumber
\end{align}
since for the IA in 2-user X channel with two antennas, each user achieves diversity gain of 2 and DoF of $\frac{4}{3}$ via IA with Alamouti coding \cite{IAAla}.
Since the dominant scale of $\textrm{Pr}\left\{\mathcal{E}_{IAA_1}\cup\mathcal{E}_{IAA_2}\right\}$ can be found in a similar way of the proof of \emph{Theorem 1}, we only sketch a proof in Appendix B, instead of a detailed one. 
\end{IEEEproof}

\begin{remark}
The reason for the DMT improvement via the on-off switched IA with Alamouti coding, in \emph{Theorem 2}, is basically  similar  to that of \emph{Theorem 1}. Moreover, the inherent merit of IA with Alamouti coding in terms of diversity yields better DMT performance than the on-off switched IA with symbol extension. 
\end{remark}

\section{Optimization and Results}
DMT of the two on-off switched IA schemes can be maximized by optimizing the time portion of IA based beamformer utilization, $a$. The corresponding optimization problems for the on-off switched IA scheme with symbol extension and the on-off switched IA with Alamouti coding are given, respectively, by
$d_{I\!A}^*(r)\!=\!\max_a\min\!\Big\{\!
\min\Big(\!
\frac{2}{a+3},\frac{1}{4a}\Big)(6-3r-2a),\min\!\Big(\frac{4}{a+3},\frac{1}{4a}\Big)(6-6r+2a),d^*_{2,2}(r)\Big\}$ and
$d_{I\!A\!A}^*(r)\!=\!\max_a\min\!\Big\{\!
\min\Big(\!
\frac{2}{a+3},\frac{1}{2a}\Big)(6-3r-2a),\min\!\Big(\frac{4}{a+3},\frac{1}{2a}\Big)(6-6r+2a),d^*_{2,2}(r)\Big\}.$

\begin{theorem}
The optimal time portion of IA utilization in the on-off switched IA scheme with symbol extension under local CSIT is determined as
\begin{eqnarray}
\label{Eq:timepIA}
    a(r)=
\left\{
\begin{array}{ccc}
    0,
    &\textrm{if }& 0<r\leq\frac{4}{5} \\
    \frac{1}{5},
    &\textrm{if }&  \frac{4}{5}<r\leq1\\
    \frac{3}{4}r,
    &\textrm{if }&  1<r\leq\frac{4}{3}\\
\end{array}.
\right.
\end{eqnarray}
\end{theorem}
\begin{IEEEproof}
Let $f_{IA_1}(a,r)=\frac{2}{a+3}(6-3r-2a)$, $f_{IA_2}(a,r)=\frac{1}{4a}(6-3r-2a)$, $f_{IA_3}(a,r)=\frac{4}{a+3}(6-6r+2a)$, and $f_{IA_4}(a,r)=\frac{1}{4a}(6-6r+2a)$.
It is readily verified that both $f_{IA_1}(a,r)$ and $f_{IA_2}(a,r)$ are decreasing functions as $a$ increases for given $r\in [0,\frac{4}{3}]$.
$f_{IA_3}(a,r)$ is an increasing function as $a$ increases for given $r\in [0,\frac{4}{3}]$ since $\frac{d}{da}f_{IA_3}(a,r)=\frac{24r}{(a+3)^2}.$
$f_{IA_4}(a,r)$ is a decreasing function as $a$ increases for given $r\in [0,1]$ and an increasing function as $a$ increases for given $r\in [1,\frac{4}{3}]$
since $\frac{d}{da}f_{IA_4}(a,r)=\frac{3(r-1)}{2a^2}.$ Therefore, we need to find the optimal time portion of  IA utilization according to the multiplexing gain region which  determines
whether $f_{IA_4}(a,r)$ is either a decreasing or an increasing function; $0 \leq r \leq 1$ and $1 \leq r \leq \frac{4}{3}$.

\begin{figure}[!t]
\centering
  \includegraphics[width=0.8\columnwidth]{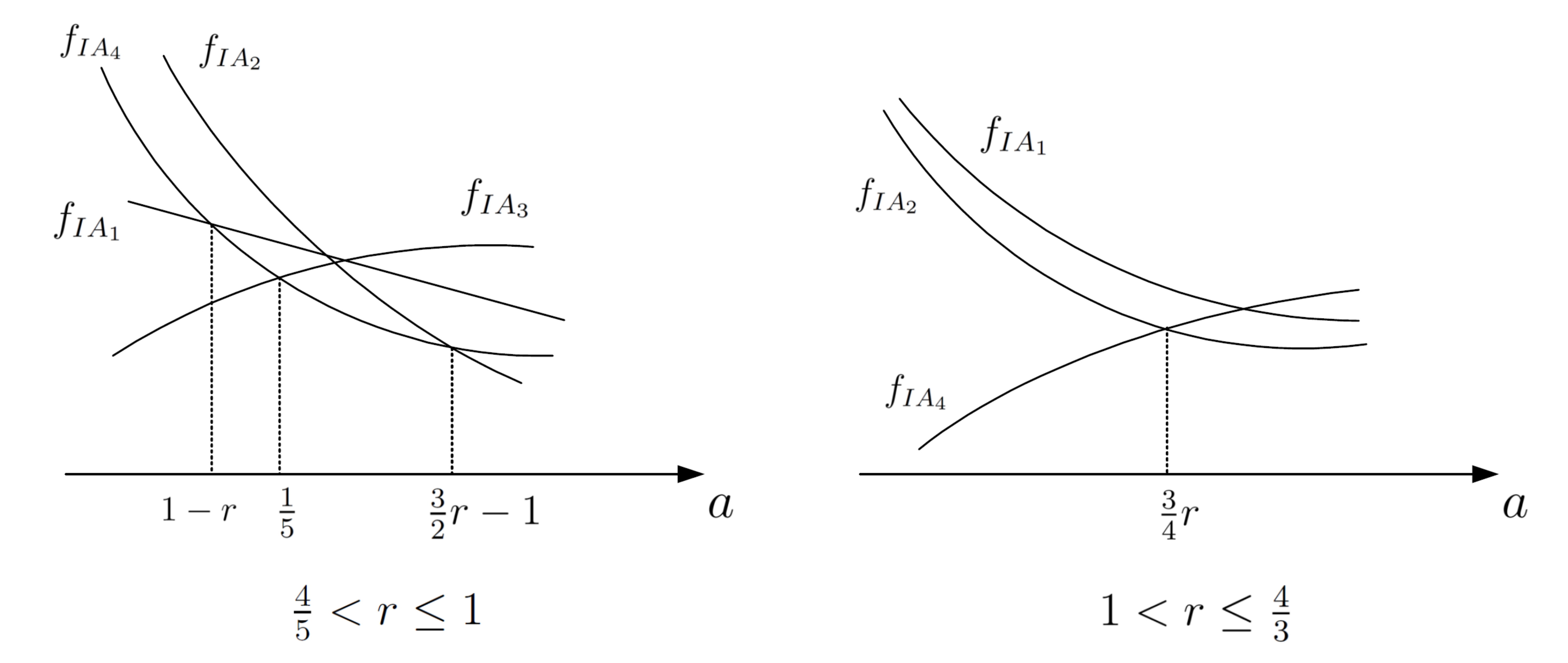}\\
  \caption{$f_{IA_i}(a,r)$ for $\frac{4}{5} \leq r \leq \frac{4}{3} $, $\forall i=1,\ldots,4$}
  \label{fig:TSi}
\end{figure}

When $0 \leq r \leq 1$, the optimal time portion of  IA utilization is 0 since the DMT of the proposed scheme with simultaneous non-unique decoding and the DMT of two-antenna point-to-point MIMO are the same
if $0\leq r \leq \frac{4}{5}$.
If $\frac{4}{5}\leq r \leq 1$, as shown in the left side of Fig. \ref{fig:TSi},
the intersection point of $f_{IA_2}(a,r)$ and $f_{IA_4}(a,r)$, that of $f_{IA_1}(a,r)$ and $f_{IA_4}(a,r)$, and that of $f_{IA_3}(a,r)$ and $f_{IA_4}(a,r)$ are
$(\frac{3}{2}r-1, f_{IA_4}(\frac{3}{2}r-1,r))$,
$(1-r, f_{IA_4}(1-r,r))$, and $(\frac{1}{5}, f_{IA_4}(\frac{1}{5},r))$, respectively. Since $\max_{a=\frac{3}{2}r-1, 1-r, \frac{1}{5}} f_{IA_4}(a,r)=f_{IA_4}(\frac{1}{5},r)$, the optimal $a(r)$ is $\frac{1}{5}$ if $\frac{4}{5}\leq r \leq 1$.

When $1 \leq r \leq \frac{4}{3}$, $f_{IA_4}(a,r)$ is less than $f_{IA_3}(a,r)$, and
 $f_{IA_2}(a,r)$ is less than $f_{IA_1}(a,r)$, as shown in the right side of Fig. \ref{fig:TSi},
Thus, if $1 \leq r \leq \frac{4}{3}$, the optimal time portion of IA utilization can be found at
the intersection point of $f_{IA_2}(a,r)$ and $f_{IA_4}(a,r)$.
Since the intersection point of $f_{IA_2}(a,r)$ and $f_{IA_4}(a,r)$ is $(\frac{3}{4}r, f_{IA_4}(\frac{3}{4}r,r))$, the optimal $a(r)$ is $\frac{3}{4}r$ if $1 \leq r \leq \frac{4}{3}$.
\end{IEEEproof}

\begin{theorem}
The optimal time portion of  IA utilization in the on-off switched IA with Alamouti coding under local CSIT is determined as
\begin{eqnarray}
\label{Eq:timepIAA}
    a(r)=
\left\{
\begin{array}{ccc}
    0,
    &\textrm{if }& 0<r\leq\frac{4}{5} \\
    \frac{3}{2}r-1,
    &\textrm{if }&  \frac{4}{5}<r\leq\frac{20}{21}\\
    \frac{3}{7},
    &\textrm{if }&  \frac{20}{21}<r\leq1\\
    \frac{3(2-r)+3\sqrt{r^2+16r-16}}{10},
    &\textrm{if }&  1<r\leq\frac{4}{3}\\
\end{array}.
\right.
\end{eqnarray}
\end{theorem}
\begin{IEEEproof}
 Let $f_{IAA_1}(a,r)=\frac{2}{a+3}(6-3r-2a)$, $f_{IAA_2}(a,r)=\frac{1}{2a}(6-3r-2a)$, $f_{IAA_3}(a,r)=\frac{4}{a+3}(6-6r+2a)$, and $f_{IAA_4}(a,r)=\frac{1}{2a}(6-6r+2a)$.
With these functions, the optimal time portion of IA utilization in the on-off switched IA with Alamouti coding  can be derived in a similar way of the proof of \emph{Theorem 3}. We skip the detailed proof due the page limit. 
\end{IEEEproof}

\begin{figure}[!t]
\centering
  \includegraphics[width=0.95\columnwidth]{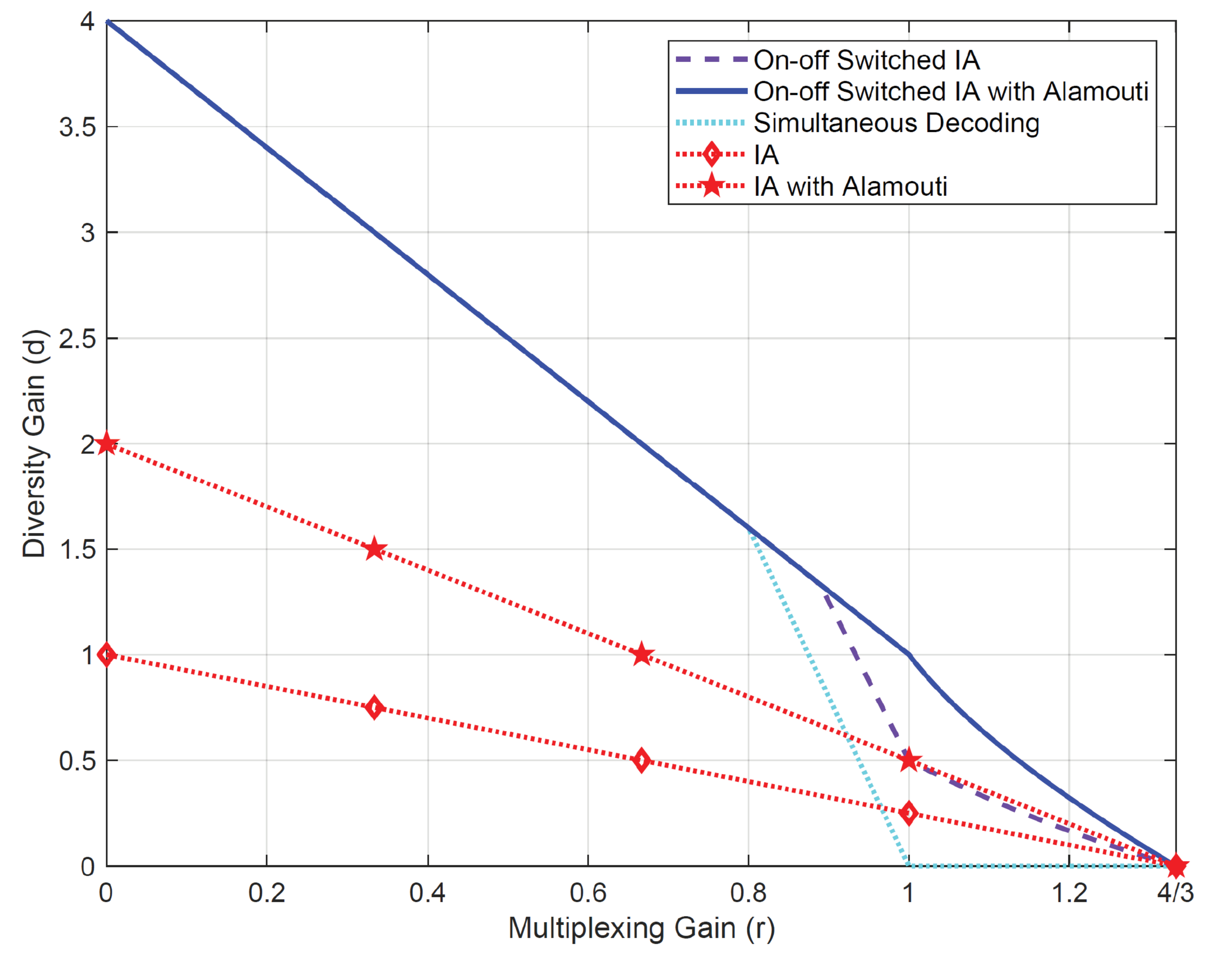}\\
  \caption{DMT of the 2-user MIMO X-channels with the on-off switched IA schemes based on IA with symbol extension and IA with Alamouti code}
  \label{fig:DMT}
\end{figure}

Fig. \ref{fig:DMT} shows DMT of  the conventional IA, the simultaneous decoding scheme without IA, the IA with Alamouti coding, the on-off switched IA with symbol extension, and the on-off switched IA with Alamouti coding in the 2-user MIMO
X-channel. Note that the simultaneous decoding scheme without IA achieves the same DMT performance of the optimal 2-user multiple access channel with two antennas \cite{MACdmt}.
The DMT of the on-off switched IA schemes 
achieves the optimal DMT of the two-antenna point-to-point MIMO in the regime where multiplexing gain is less than about 0.9.
The on-off switched IA with symbol extension outperforms the conventional
IA with Alamouti coding (without switching) in terms of DMT for $0\leq r \leq 1$.
In particular, the DMT of the on-off switched IA with Alamouti coding not only surpasses  the conventional IA with Alamouti coding (without switching) but also approaches to the linear function connecting
two points: maximum diversity gain and multiplexing gain per user, i.e., (0,4) and (4/3,0), respectively.

\section{Discussion: Extensions to Arbitrary $M$}
The IA based Alamouti coding scheme for $M=2$ was extended to more than two antennas in \cite{IASTBC3}, \cite{IASTBC4} and \cite{IASTBCM}, which proposed  interference aligned space-time transmission schemes  when the number of antenna at each node, $M$, is 3, 4, and arbitrary, respectively. They constructed the transmit codewords tailored to the number of antennas, $M$, based on symbol extension, and showed that the diversity gain of $M$ can be achieved. Meanwhile, the optimal  DoF region of a two-user X-channel CSIT was characterized under asymmetric output feedback and delayed when transmitters and receivers are equipped with arbitrary number of antennas at each \cite{Editor},\cite{Editor2}. Therefore, for more than two antennas, i.e., $M\geq3$,  on-off switching can  be readily applied to the interference aligned space time transmission schemes, as it is applied to the IA based Alamouti coding for $M=2$.  The optimization of the switching portion, $a$, could be non-trivially complicated but the fundamental mechanism is retained even for arbitrary $M$ whereby the proposed on-off switched IA scheme will be still effective in improving the DMT. More specifically, if on-off switching is applied to the interference aligned space-time transmission of \cite{IASTBCM} for an arbitrary number of antennas, $M$,  the full diversity gain of $M^2$ and the optimal multiplexing gain of  $\frac{4M}{3}$, as well as DMT improvement in intermediate multiplexing gain regime,  can be achieved. However, the exact derivation of the DMT by optimizing the switching portion, $a$, is non-trivially complicated to be addressed in this paper.

\section{Conclusion}
We proposed the on-off switched IA schemes to verify that  the intermittent utilization of IA can improve DMT in interference channels.  For switching on/off, either the IA with symbol extension or the IA with Alamouti coding is used in the proposed on-off switched IA. We derived DMT of the two proposed schemes in closed form  in the 2-user X-channel with two antennas.
Both of the proposed scheme were shown to achieve diversity gain of 4 and DoF per user of $\frac{4}{3}$ (i.e., sum DoF of $\frac{8}{3}$), if the time portion of IA utilization is optimized. The optimized on-off switched IA scheme with symbol extension was shown to outperform the conventional IA with Alamouti coding (without switching) for $0 \leq r \leq 1$, although Alamouti coding is not exploited.
The optimized on-off switched IA with Alamouti coding scheme,
to the best our knowledge, surpasses any other existing schemes in the 2-user X-channel with two antennas, and approaches to
the linear function connecting maximum diversity gain and DoF per user, (0,4) and $(\frac{4}{3},0)$, respectively, i.e., $d(r)=4-3r$.

\appendices
\section{Finding the Dominant Scale of $\textrm{Pr}\{\mathcal{E}_{IA_1}\}$ and $\textrm{Pr}\{\mathcal{E}_{IA_2}\}$}

From (\ref{Eq:ets1}), we consider the cases of $v_{11}^{[11]}+v_{22}^{[11]}<v_{12}^{[11]}+v_{21}^{[11]}$ and $v_{11}^{[11]}+v_{22}^{[11]}\geq v_{12}^{[11]}+v_{21}^{[11]}$
in order to find the dominant scale of $\textrm{Pr}\{\mathcal{E}_{IA_1}\}$.

If $v_{11}^{[11]}+v_{22}^{[11]}<v_{12}^{[11]}+v_{21}^{[11]}$,
\begin{align}
\label{Eq:eqOut1}
\textrm{Pr}\{\mathcal{E}_{IA_1}\}\!\!&=\!\textrm{Pr}\Bigg\{\!\frac{4a(1\!-\!v_{11}^{[11]})}{3(1-a)}\!+\!2\!-\!v_{11}^{[11]}\!-\!v_{22}^{[11]}\!\!<\!\!\frac{r}{1-a}\!\Bigg\}
\nonumber\\\!\!&\doteq\!\textrm{Pr}\Bigg\{\!(a\!+\!3)v_{11}^{[11]}\!+\!3(\!1\!-\!a)v_{22}^{[11]}\!\!>\!6\!-\!3r\!-\!2a\!\Bigg\}.\nonumber
\end{align}
Let us define an outage set, $\mathcal{O}_1=\{(v_{11}^{[11]},\ldots,v_{22}^{[11]}) |
(a+3)v_{11}^{[11]}+3(1-a)v_{22}^{[11]}>6-3r-2a\}$, in terms of the associated exponential
variables. Then, from (\ref{eq:expresult}),
\begin{align}
\textrm{Pr}\{\mathcal{E}_{IA_{1}}\}&\doteq\rho^{-\inf_{(v_{11}^{[11]},\ldots,v_{22}^{[11]})\in \mathcal{O}_1^+}\sum_{m,n=,1,2}v_{mn}^{[11]}}
\nonumber\\&\stackrel{(\textrm{a})}\doteq\rho^{-\inf_{(v_{11}^{[11]},\ldots,v_{22}^{[11]})\in \mathcal{O}_1^+}2(v_{11}^{[11]}+v_{22}^{[11]})},
\end{align}
where (a) follows from $v_{11}^{[11]}+v_{22}^{[11]}<v_{12}^{[11]}+v_{21}^{[11]}$.

\begin{figure}[!t]
\centering
  \includegraphics[width=1\columnwidth]{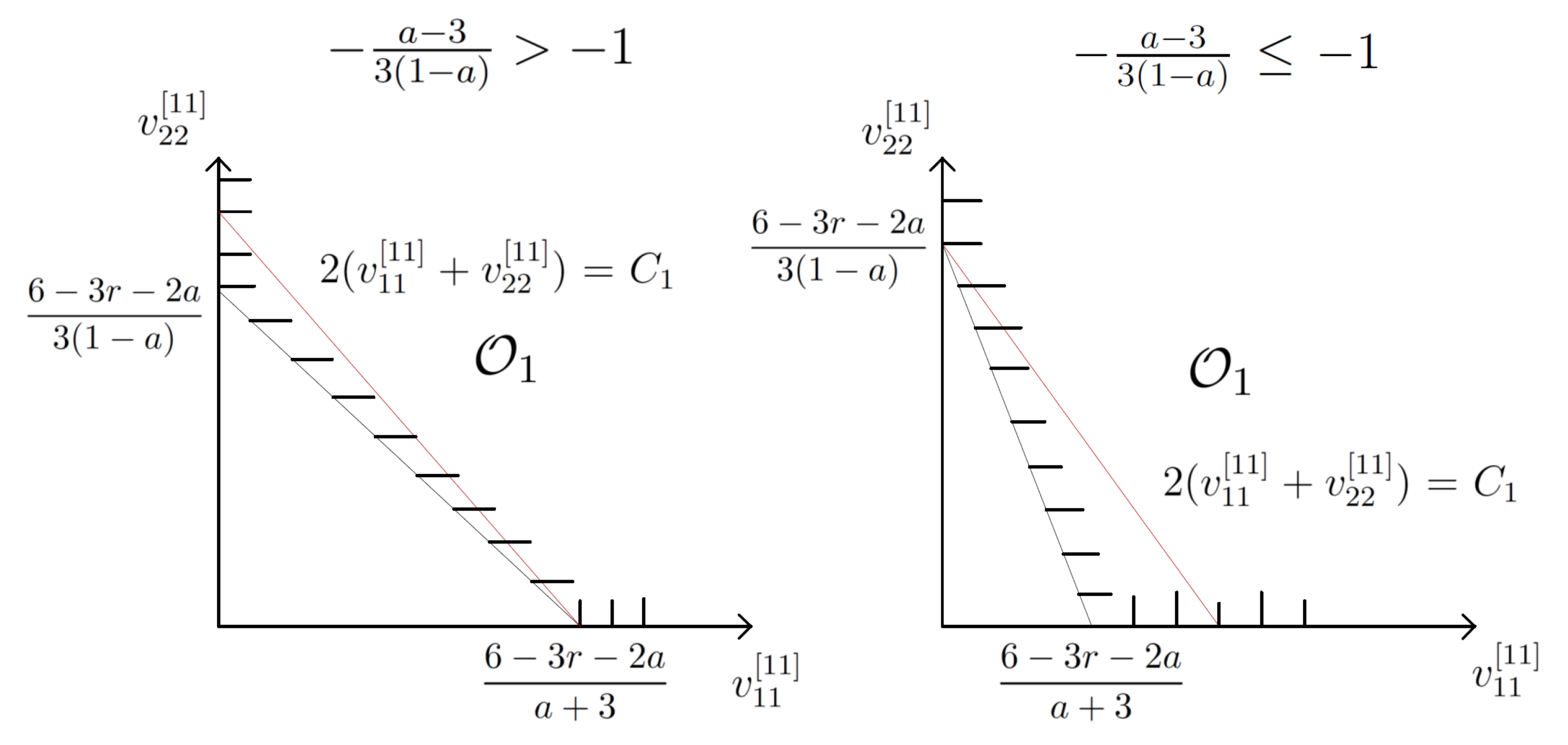}\\
  \caption{The description of $\mathcal{O}_1^+$ and $\inf_{(v_{11}^{[11]},\ldots,v_{22}^{[11]})\in \mathcal{O}_1^+}2(v_{11}^{[11]}+v_{22}^{[11]})$ depending on $-\frac{a-3}{3(1-a)}$.}
  \label{fig:O1}
\end{figure}

For simple notation, let us define $2(v_{11}^{[11]}+v_{22}^{[11]})$ as $C_1$ (i.e., $2(v_{11}^{[11]}+v_{22}^{[11]})=C_1$).  Then, as illustrated
in Fig. \ref{fig:O1},
$\min
C_1$ varies with $-\frac{a-3}{3(1-a)}$ that is the slope of $(a+3)v_{11}^{[11]}+3(1-a)v_{22}^{[11]}>6-3r-2a$ in $\mathcal{O}_1$.
If $-\frac{a-3}{3(1-a)}> -1$, $C_1$ is minimized when it is on the $v_{11}^{[11]}$ axis and thus
$\inf_{(v_{11}^{[11]},\ldots,v_{22}^{[11]})\in \mathcal{O}_1^+}2(v_{11}^{[11]}+v_{22}^{[11]})=2v_{22}^{[11]}=\frac{2}{3(1-a)}(6-3r-2a)$. Otherwise,
if $-\frac{a-3}{3(1-a)}\leq -1$,
since $C_1$ is minimized when it is on the $v_{22}^{[11]}$ axis,
$\inf_{(v_{11}^{[11]},\ldots,v_{22}^{[11]})\in \mathcal{O}_1^+}2(v_{11}^{[11]}+v_{22}^{[11]})=2v_{11}^{[11]}=\frac{2}{a+3}(6-3r-2a)$.
Therefore, if $v_{11}^{[11]}+v_{22}^{[11]}<v_{12}^{[11]}+v_{21}^{[11]}$,
$\textrm{Pr}\{\mathcal{E}_{IA_{1}}\}\doteq \rho^{-\min(\frac{1}{a+3},\frac{1}{3(1-a)})2(6-3r-2a)}.$

\begin{figure}[t!]
\centering
  \includegraphics[width=1\columnwidth]{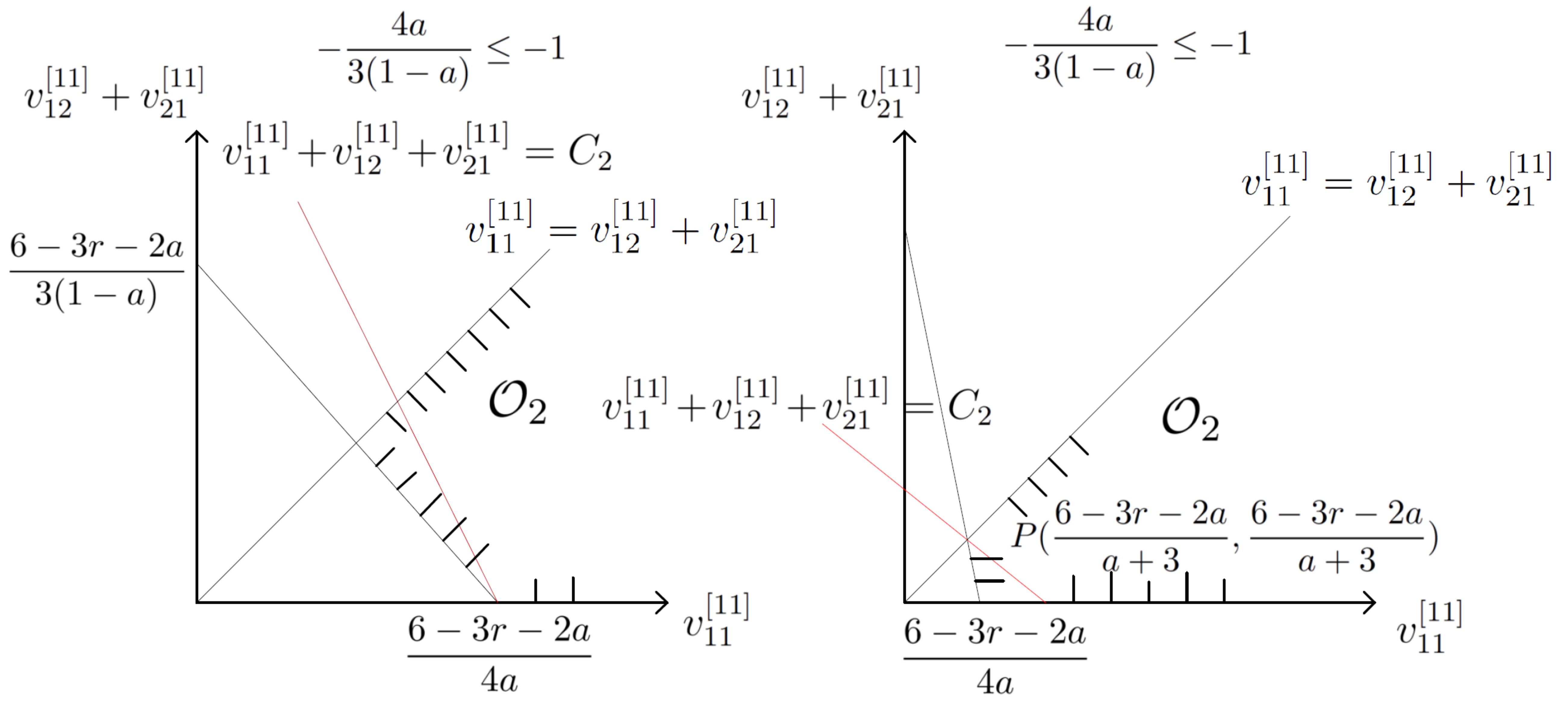}\\
  \caption{The description of $\mathcal{O}_2^+$ and $\inf_{(v_{11}^{[11]},\ldots,v_{22}^{[11]})\in \mathcal{O}_2^+}v_{11}^{[11]}+v_{12}^{[11]}+v_{21}^{[11]}$ depending on $-\frac{4a}{3(1-a)}$.}
  \label{fig:O2}
\end{figure}

If $v_{11}^{[11]}+v_{22}^{[11]}\geq v_{12}^{[11]}+v_{21}^{[11]}$,
$\textrm{Pr}\{\mathcal{E}_{IA_1}\}\!\doteq\!\textrm{Pr}\Big\{\!4a(1-v_{11}^{[11]})+\!3(\!1\!-\!a\!)(2\!-\!v_{12}^{[11]}\!-\!v_{21}^{[11]})<3r\Big\}\doteq\textrm{Pr}\Big\{\!4av_{11}^{[11]}+3(1-a)(v_{12}^{[11]}+v_{21}^{[11]})\!\!>\!6\!-\!3r\!-\!2a\!\Big\}.$
Let us define an outage set, $\mathcal{O}_2=\{(v_{11}^{[11]},\ldots,v_{22}^{[11]}) |
4av_{11}^{[11]}+3(1-a)(v_{12}^{[11]}+v_{21}^{[11]})>6-3r-2a\}$.
In a similar way to the case when $v_{11}^{[11]}+v_{22}^{[11]}<v_{12}^{[11]}+v_{21}^{[11]}$,
\begin{align}
\textrm{Pr}\{\mathcal{E}_{IA_{1}}\}&\doteq\rho^{-\inf_{(v_{11}^{[11]},\ldots,v_{22}^{[11]})\in \mathcal{O}_2^+}\sum_{m,n=,1,2}v_{mn}^{[11]}}
\nonumber\\&\stackrel{(\textrm{b})}\doteq\rho^{-\inf_{(v_{11}^{[11]},\ldots,v_{22}^{[11]})\in \mathcal{O}_2^+}(v_{11}^{[11]}+v_{12}^{[11]}+v_{21}^{[11]})},
\end{align}
where (b) follows from $v_{22}^{[11]}>0$.  When (b) holds, the condition $v_{11}^{[11]}+v_{22}^{[11]}\geq v_{12}^{[11]}+v_{21}^{[11]}$ is reduced to
$v_{11}^{[11]}\geq v_{12}^{[11]}+v_{21}^{[11]}$ since outage set $\mathcal{O}_2$ is not dependent on $v_{22}^{[11]}$ and $\inf_{v_{22}^{[11]}>0} v_{22}^{[11]}=0$.
Let $v_{11}^{[11]}+v_{12}^{[11]}+v_{21}^{[11]}=C_2$. Then, using a plane with $v_{11}^{[11]}$ and $v_{12}^{[11]}+v_{21}^{[11]}$ axes in
Fig. \ref{fig:O2}, we can obtain the dominant scale of $\textrm{Pr}\{\mathcal{E}_{IA_{1}}\}$,  when $v_{11}^{[11]}+v_{22}^{[11]}\geq v_{12}^{[11]}+v_{21}^{[11]}$, as
$\textrm{Pr}\{\mathcal{E}_{IA_{1}}\}\doteq \rho^{-\min(\frac{2}{a+3}, \frac{1}{4a})(6-3r-2a)}.$

Therefore, combining the two cases when $v_{11}^{[11]}+v_{22}^{[11]}<v_{12}^{[11]}+v_{21}^{[11]}$ and when $v_{11}^{[11]}+v_{22}^{[11]}\geq v_{12}^{[11]}+v_{21}^{[11]}$, we obtain the dominant scale of $\textrm{Pr}\{\mathcal{E}_{IA_{1}}\}$ as
\begin{align}
\textrm{Pr}\{\mathcal{E}_{IA_{1}}\}&\doteq \rho^{-\min(\frac{2}{3(1-a)},\frac{2}{a+3}, \frac{1}{4a})(6-3r-2a)}
\nonumber\\&\doteq\rho^{-\min(\frac{2}{a+3},\frac{1}{4a})(6-3r-2a)},
\end{align}
since $\frac{2}{a+3} \leq \frac{2}{3(1-a)}$ for $0 \leq a \leq 1$.

Now, we calculate the dominant scale of $\textrm{Pr}\{\mathcal{E}_{IA_2}\}$.
If $v_{1k_1}^{[1l_1]}=v_{11}^{[11]}$, the number of independent random variables involved in (\ref{Eq:ets2}) is 3; otherwise, i.e.,  $v_{1k_1}^{[1l_1]}\neq v_{11}^{[11]}$, the number of independent random variables involved in (\ref{Eq:ets2}) is 4.

\begin{figure}[t!]
\centering
  \includegraphics[width=0.8\columnwidth]{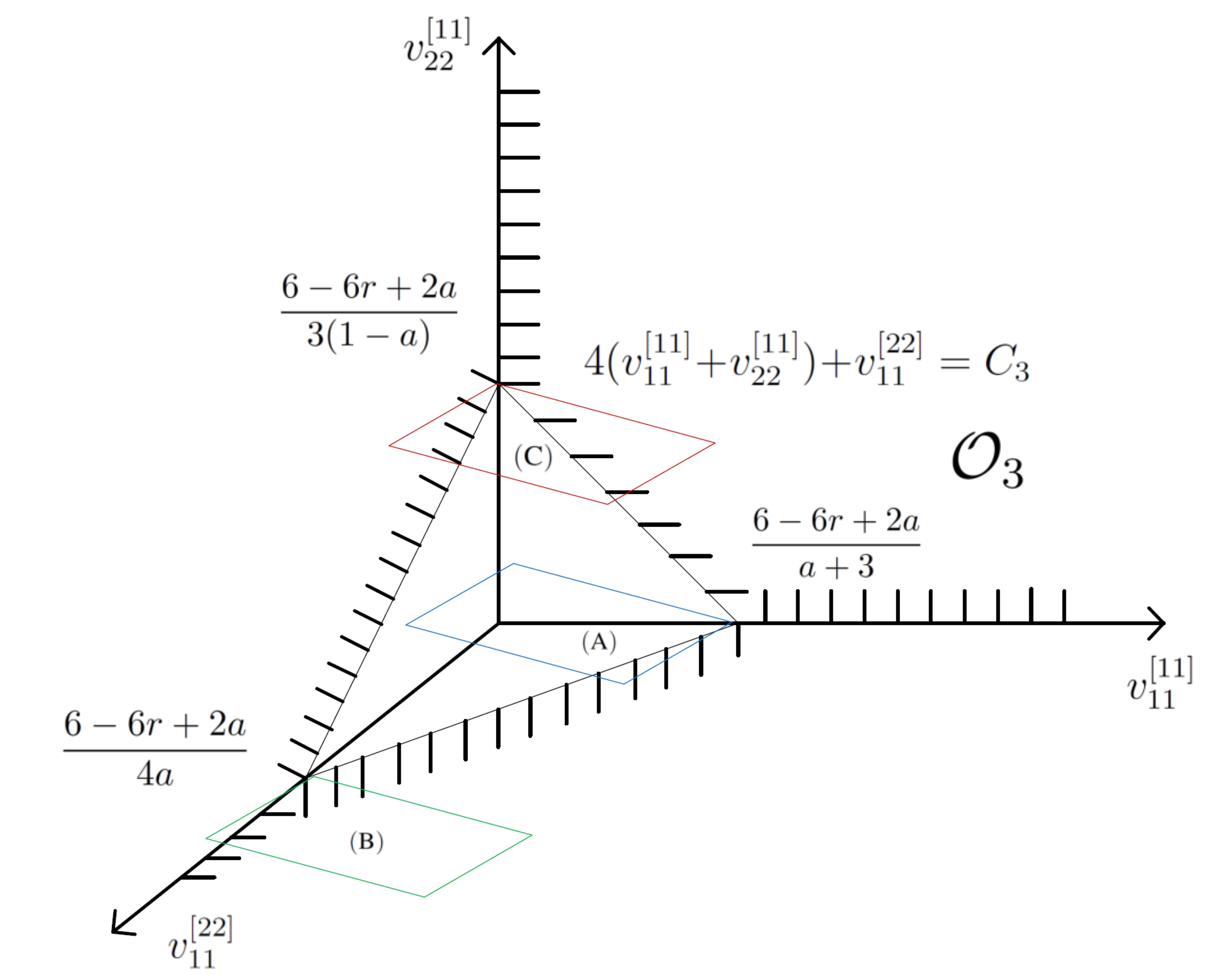}\\
  \caption{The description of $\mathcal{O}_3^+$ and $\inf_{(v_{11}^{[11]},\ldots,v_{22}^{[11]})\in \mathcal{O}_3^+} 4(v_{11}^{[11]}+v_{22}^{[11]})+v_{11}^{[22]}$ depending on $(a+3)v_{11}^{[11]}+3(1-a)v_{22}^{[11]}+4av_{11}^{[22]}=6-6r+2a$.}
  \label{fig:O3}
\end{figure}
Firstly, we consider the case when $v_{1k_1}^{[1l_1]}=v_{11}^{[11]}$. In this case.
$v_{2k_2}^{[1l_2]}$ can be $v_{21}^{[12]}$, $v_{22}^{[11]}$, or $v_{22}^{[12]}$.
Since $v_{21}^{[12]}$, $v_{22}^{[11]}$, and $v_{22}^{[12]}$ are \emph{i.i.d.}, and dominant scale of $\textrm{Pr}\{\mathcal{E}_{IA_2}\}$ is the same whatever $v_{2k_2}^{[1l_2]}$ is among the three possible candidates,
without loss of generality, we consider only one of the candidates, $v_{2k_2}^{[1l_2]}=v_{22}^{[11]}$. Then, since
\begin{align}
\label{Eq:eqcon}
\min_{l_1,l_2,k_1,k_2\in \{1,2\}\bigcap(l_1,k_1)\neq(l_2,k_2)}v_{1k_1}^{[1l_1]}+v_{2k_2}^{[1l_2]}=v_{11}^{[11]}+v_{22}^{[11]},
\end{align} the dominant scale of the outage probability of $\mathcal{E}_{IA_2}$ becomes
\begin{align}
\textrm{Pr}\{\mathcal{E}_{IA_2}\}&\doteq\textrm{Pr}\Big\{\rho^{\frac{4a}{3(1-a)}(2-v_{11}^{[11]}-v_{11}^{[22]})+2-v_{11}^{[11]}-v_{22}^{[11]}}<\rho^{\frac{2r}{1-a}} \Big\}\nonumber\\
&\doteq\textrm{Pr}\Big\{(a+3)v_{11}^{[11]}+3(1-a)v_{22}^{[11]}+4av_{11}^{[22]}
\nonumber\\&~~~~~~>6-6r+2a \Big\}.
\end{align}
Let us define an outage set, $\mathcal{O}_3=\{(v_{11}^{[11]},\ldots, v_{22}^{[12]}, v_{11}^{[22]}) |
(a+3)v_{11}^{[11]}+3(1-a)v_{22}^{[11]}+4av_{11}^{[22]}>6-6r+2a\}$, in terms of the associated exponential
variables. Then, from (\ref{eq:expresult}),
\begin{align}
\textrm{Pr}\{\mathcal{E}_{IA_{2}}\}&\doteq\rho^{-\inf_{(v_{11}^{[11]},\ldots, v_{22}^{[12]}, v_{11}^{[22]})\in \mathcal{O}_3^+}\sum_{p,q,r = 1,2}v_{1p}^{[qr]}+v_{11}^{[22]}}\nonumber\\
&\stackrel{(\textrm{c})}\doteq\rho^{-\inf_{(v_{11}^{[11]},\ldots, v_{22}^{[12]}, v_{11}^{[22]})\in \mathcal{O}_3^+}4(v_{11}^{[11]}+v_{22}^{[11]})+v_{11}^{[22]}},\nonumber
\end{align}
where (c) follows from (\ref{Eq:eqcon}).
Let $4(v_{11}^{[11]}+v_{22}^{[11]})+v_{11}^{[22]}=C_3$.
Then, as illustrated in Fig. \ref{fig:O3},
since both $4(v_{11}^{[11]}+v_{22}^{[11]})+v_{11}^{[22]}=C_3$ and $(a+3)v_{11}^{[11]}+3(1-a)v_{22}^{[11]}+4av_{11}^{[22]}=6-6r+2a$ are planes in the three-dimensional space,
$\inf C_3$ is found at a point on $v_{11}^{[11]}$, $v_{11}^{[22]}$, or $v_{22}^{[11]}$ axis, which correspond to cases (A), (B), and (C) in Fig. \ref{fig:O3}; For case (A), $\inf C_3$ is found at $(0,\frac{6-6r+2a}{a+3},0)$ on $v_{11}^{[11]}$ axis. Hence, the dominant scale of $\textrm{Pr}\{\mathcal{E}_{IA_{2}}\}$ is represented as
$\rho^{-\inf_{(v_{11}^{[11]},\ldots, v_{22}^{[12]}, v_{11}^{[22]})\in \mathcal{O}_3^+}4v_{11}^{[11]}}\doteq\rho^{-\frac{4(6-6r+2a)}{a+3}}.$
For case (B), $\inf C_3$ is found at $(\frac{6-6r+2a}{4a},0,0)$ on $v_{11}^{[22]}$ axis, and the dominant scale of $\textrm{Pr}\{\mathcal{E}_{IA_{2}}\}$ is represented as
$\rho^{-\inf_{(v_{11}^{[11]},\ldots, v_{22}^{[12]}, v_{11}^{[22]})\in \mathcal{O}_3^+}v_{11}^{[22]}}\doteq\rho^{-\frac{(6-6r+2a)}{4a}}.$
For case (C), $\inf C_3$ is found at $(0,0,\frac{6-6r+2a}{3(1-a)})$ on $v_{22}^{[11]}$ axis, and similarly, the dominant scale of $\textrm{Pr}\{\mathcal{E}_{IA_{2}}\}$ is represented as
$\rho^{-\inf_{(v_{11}^{[11]},\ldots, v_{22}^{[12]}, v_{11}^{[22]})\in \mathcal{O}_3^+}4v_{22}^{[11]}}\doteq\rho^{-\frac{4(6-6r+2a)}{3(1-a)}}.$
\begin{align}
\textrm{Pr}\{\mathcal{E}_{IA_{2}}\}&\doteq \rho^{-\min(\frac{4}{3(1-a)},\frac{4}{a+3},\frac{1}{4a})(6-6r+2a)}\nonumber\\
&\doteq \rho^{-\min(\frac{4}{a+3},\frac{1}{4a})(6-6r+2a)},
\end{align}
since $\frac{4}{3(1-a)}(6-6r+2a)\geq\frac{4}{a+3}(6-6r+2a), \forall r, 0\leq a \leq 1 $.

Secondly, we consider the case when $v_{1k_1}^{[1l_1]}\neq v_{11}^{[11]}$.
If $v_{1k_1}^{[1l_1]}\neq v_{11}^{[11]}$, the number of possible pairs of $(v_{1k_1}^{[1l_1]},v_{2k_2}^{[1l_2]})$ resulting in the same value of $\min_{l_1,l_2,k_1,k_2\in \{1,2\}\bigcap(l_1,k_1)\neq(l_2,k_2)}v_{1k_1}^{[1l_1]}+v_{2k_2}^{[1l_2]}$ is 9. Therefore, similar to the case when $v_{1k_1}^{[1l_1]}= v_{11}^{[11]}$,
we consider only one of the possible 9 candidates that
\begin{align}
\label{Eq:eqcon2}
\min_{l_1,l_2,k_1,k_2\in \{1,2\}\bigcap(l_1,k_1)\neq(l_2,k_2)}v_{1k_1}^{[1l_1]}+v_{2k_2}^{[1l_2]}=v_{11}^{[12]}+v_{22}^{[11]}.
\end{align}
With (\ref{Eq:eqcon2}), the dominant scale of the outage probability of $\mathcal{E}_{IA_2}$ becomes
\begin{align}
\textrm{Pr}\{\mathcal{E}_{IA_2}\}&\doteq\textrm{Pr}\Big\{\rho^{\frac{4a}{3(1-a)}(2-v_{11}^{[11]}-v_{11}^{[22]})+2-v_{11}^{[12]}-v_{22}^{[11]}}<\rho^{\frac{2r}{1-a}} \Big\}\nonumber\\
&\doteq\textrm{Pr}\Big\{3(1-a)(v_{11}^{[12]}+v_{22}^{[11]})+4a(v_{11}^{[11]}+v_{11}^{[22]})\nonumber\\
&~~~~~>6-6r+2a \Big\}.
\end{align}
Let us define an outage set, $\mathcal{O}_4=\{(v_{11}^{[11]},\ldots, v_{22}^{[12]}, v_{11}^{[22]}) |
3(1-a)(v_{11}^{[12]}+v_{22}^{[11]})+4a(v_{11}^{[11]}+v_{11}^{[22]})>6-6r+2a\}$, in terms of the associated exponential
variables. Then, from (\ref{eq:expresult}),
\begin{align}
\textrm{Pr}\{\mathcal{E}_{IA_{2}}\}\!&\doteq\rho^{-\inf_{(v_{11}^{[11]},\ldots, v_{22}^{[12]}, v_{11}^{[22]})\in \mathcal{O}_4^+}\sum_{p,q,r = 1,2} v_{1p}^{[qr]}+v_{11}^{[22]}}\\
&\stackrel{(\textrm{d})}\doteq\!\rho^{\!\!\!-\inf_{(v_{11}^{[11]},\!\ldots, v_{22}^{[12]}\!, v_{11}^{[22]})\in \mathcal{O}_4^+}\!\! 3(v_{11}^{[12]}\!+v_{22}^{[11]})\!+v_{11}^{[11]}\!\!+v_{12}^{[12]}\!\!+v_{11}^{[22]}}\\
&\stackrel{(\textrm{e})}\doteq\rho^{\!\!-\inf_{(v_{11}^{[11]},\ldots, v_{22}^{[12]}, v_{11}^{[22]})\in \mathcal{O}_4^+} 3(v_{11}^{[12]}+v_{22}^{[11]})+v_{11}^{[11]}+v_{11}^{[22]}},
\end{align}
where (d) follows from (\ref{Eq:eqcon2}) that $v_{11}^{[12]}+v_{22}^{[11]}\leq v_{12}^{[11]}+v_{21}^{[12]}$, $v_{21}^{[11]}+v_{22}^{[12]}$, and (e) follows from $\inf_{v_{12}^{[12]}>0}v_{12}^{[12]}=0$.  Since $\inf_{v_{12}^{[12]}>0}v_{12}^{[12]}=0$, the condition that $v_{11}^{[12]}+v_{22}^{[11]}\leq v_{11}^{[11]}+v_{12}^{[12]}$ from  (\ref{Eq:eqcon2})  reduces to $v_{11}^{[12]}+v_{22}^{[11]}\leq v_{11}^{[11]}$ in finding the dominant scale of $\textrm{Pr}\{\mathcal{E}_{IA_2}\}$.

\begin{figure}[t!]
\centering
  \includegraphics[width=0.8\columnwidth]{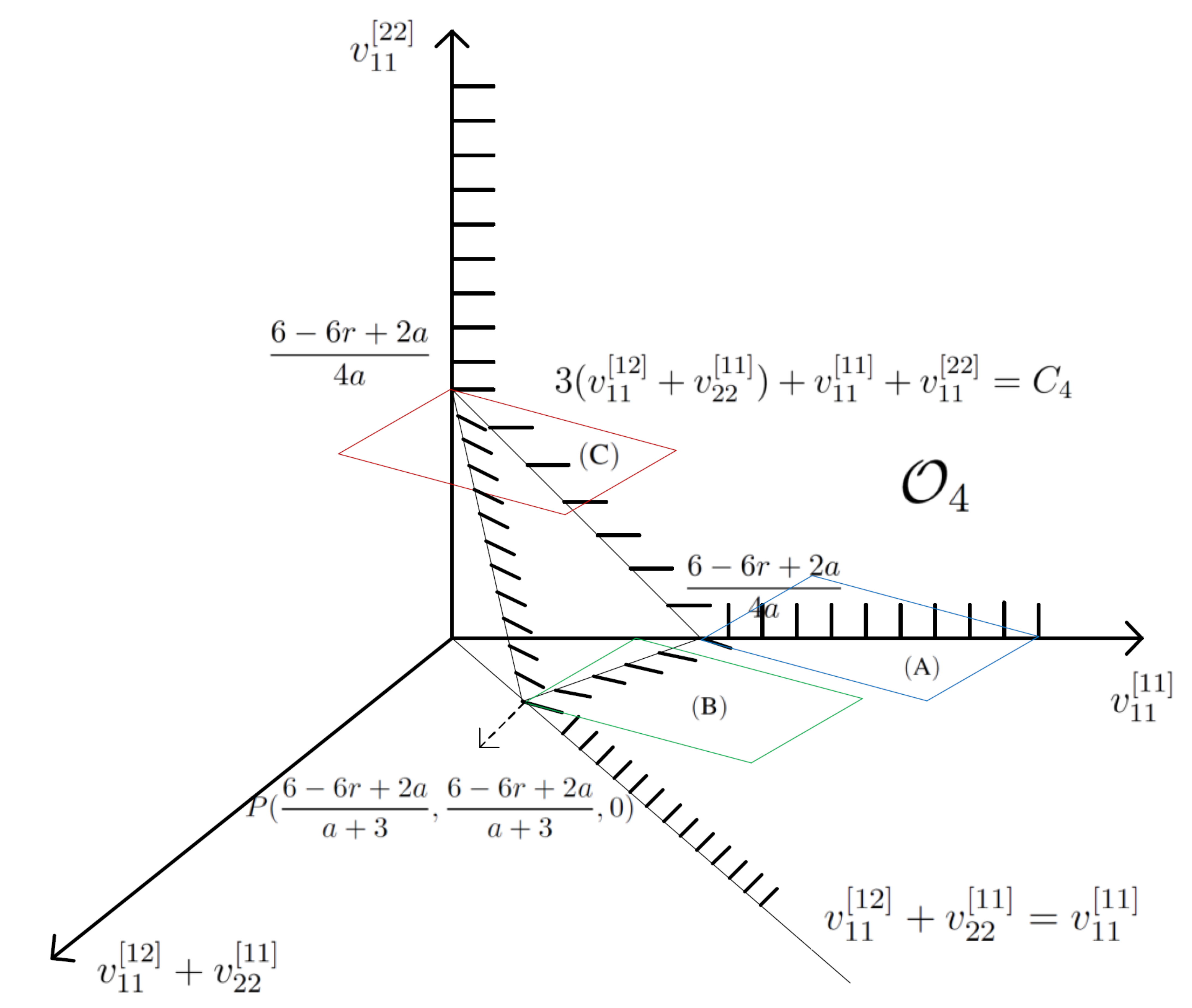}\\
  \caption{The description of $\mathcal{O}_4^+$ and $\inf_{(v_{11}^{[11]},\ldots, v_{22}^{[12]}, v_{11}^{[22]})\in \mathcal{O}_4^+} 3(v_{11}^{[12]}+v_{22}^{[11]})+v_{11}^{[11]}+v_{11}^{[22]}$ depending on $3(1-a)(v_{11}^{[12]}+v_{22}^{[11]})+4a(v_{11}^{[11]}+v_{11}^{[22]})=6-6r+2a$ under $v_{11}^{[12]}+v_{22}^{[11]}<v_{11}^{[11]}$.}
  \label{fig:O4}
\end{figure}

Let $3(v_{11}^{[12]}+v_{22}^{[11]})+v_{11}^{[11]}+v_{11}^{[22]}=C_4$. Then, as illustrated in  Fig. \ref{fig:O4}, both $3(v_{11}^{[12]}+v_{22}^{[11]})+v_{11}^{[11]}+v_{11}^{[22]}=C_4$ and $3(1-a)(v_{11}^{[12]}+v_{22}^{[11]})+4a(v_{11}^{[11]}+v_{11}^{[22]})=6-6r+2a$ are planes in the three-dimensional space. Under $v_{11}^{[12]}+v_{22}^{[11]}\leq v_{11}^{[11]}$, $\inf C_4$ can be found at  $(0,\frac{6-6r+2a}{4a},0)$, $(\frac{6-6r+2a}{a+3},\frac{6-6r+2a}{a+3},0)$, or $(0,0,\frac{6-6r+2a}{4a})$, which correspond to cases (A), (B), and (C) in Fig. \ref{fig:O4}, respectively; For case (A), $\inf C_4$ is found at $(0,\frac{6-6r+2a}{4a},0)$ on $v_{11}^{[11]}$ axis and thus the dominant scale of $\textrm{Pr}\{\mathcal{E}_{IA_{2}}\}$ is represented as
$\rho^{-\inf_{(v_{11}^{[11]},\ldots, v_{22}^{[12]}, v_{11}^{[22]})\in \mathcal{O}_4^+}v_{11}^{[11]}}
\doteq\rho^{-\frac{6-6r+2a}{4a}}.$
For case (B), $\inf C_4$ is found at $(\frac{6-6r+2a}{a+3},\frac{6-6r+2a}{a+3},0)$ on $v_{11}^{[11]}=v_{11}^{[12]}+v_{22}^{[11]}$, and the dominant scale of $\textrm{Pr}\{\mathcal{E}_{IA_{2}}\}$ is represented as
$\rho^{-\inf_{(v_{11}^{[11]},\ldots, v_{22}^{[12]}, v_{11}^{[22]})\in \mathcal{O}_4^+}3(v_{11}^{[12]}+v_{22}^{[11]})+v_{11}^{[11]}}\doteq\rho^{-\frac{4(6-6r+2a)}{a+3}}.$
For case (C), $\inf C_4$ is found at $(0,0,\frac{6-6r+2a}{4a})$ on $v_{11}^{[22]}$ axis, and similarly, the dominant scale of $\textrm{Pr}\{\mathcal{E}_{IA_{2}}\}$ is represented as
$\textrm{Pr}\{\mathcal{E}_{IA_{2}}\}\doteq\rho^{-\inf_{(v_{11}^{[11]},\ldots, v_{22}^{[12]}, v_{11}^{[22]})\in \mathcal{O}_4^+}v_{11}^{[22]}}\doteq\rho^{-\frac{(6-6r+2a)}{4a}}.$ Consequently, combining the three cases,
\begin{align}
\textrm{Pr}\{\mathcal{E}_{IA_{2}}\}&\doteq \rho^{-\min(\frac{4}{a+3},\frac{1}{4a})(6-6r+2a)}.
\end{align}


Finally, the diversity gain of the on-off switched IA scheme with symbol extension is determined by the dominant scale of $\textrm{Pr}\{\mathcal{E}_{IA_{1}}\}$ and $\textrm{Pr}\{\mathcal{E}_{IA_{2}}\}$, which is also bounded by the diversity gain of the point-to-point 2 $\times$ 2 MIMO case given as
$d(r)=\inf_{\boldsymbol{\alpha}\in\mathcal{A}}\sum_{i=1}^{2}(2i-1)\alpha_i$
where $\mathcal{A}=\left\{\:\boldsymbol{\alpha}\in\mathbb{R}^{2}\:\Big|\quad \alpha_1\geq \alpha_2\geq0,\text{ and }\sum_i(1-\alpha_i)^+<r\:\right\}$
and $\alpha_i$ is the exponential order of eigenvalue $\lambda_i$ for $i \in \{1,2\}$ \cite{dmt}. 
Therefore, the diversity gain  is determined as the minimum of $d^*_{2,2}(r)$ and the dominant scale of $\textrm{Pr}\{\mathcal{E}_{IA_1}\}$ and $\textrm{Pr}\{\mathcal{E}_{IA_2}\}$.


\section{Finding the Dominant Scale of $\textrm{Pr}\{\mathcal{E}_{IAA_1}\}$ and $\textrm{Pr}\{\mathcal{E}_{IAA_2}\}$}

Because all channels of two symmetric users are \emph{i.i.d.}, indices $i, j,$  and $k$ do not affect the dominant scale of $\textrm{Pr}\{\mathcal{E}_{IAA}\}$.
For that reason,  considering only the case of $i=j=k=1$  suffices for simple notation.

For $\mathcal{E}_{IAA_1}$, the outage probability is represented as
\begin{align}
\textrm{Pr}\{\mathcal{E}_{IAA_1}\}&\!\doteq\!\textrm{Pr}\Big\{\frac{4}{3}a\log\left(1\!+\!\rho\left(\left|h_{11}^{[11]}\right|^2\!\!\!+\!\left|h_{11}^{[12]}\right|^2\right)\right)\!\!\nonumber\\
&+\!\!(1\!-\!a)\log\det\left(\mathbf{I}\!+\rho\mathbf{H}^{[11]}\mathbf{H}^{[11]\dagger}\right)
\!<\!r\log\rho \Big\}\nonumber\\
&\doteq\textrm{Pr}\Big\{\log\left(\rho^{1-v_{11}^{[11]}}+\rho^{1-v_{11}^{[12]}}\right)^{\frac{4a}{3(1-a)}}\nonumber\\
&+\log(\rho^{2-v_{11}^{[11]}-v_{22}^{[11]}}\!\!+\!\rho^{2-v_{12}^{[11]}-v_{21}^{[11]}})\!<\!\log\rho^{\frac{r}{1-a}} \!\Big\}\nonumber\\
&\stackrel{(\textrm{f})}\leq\textrm{Pr}\Big\{\log\left(\rho^{\frac{4a}{3(1-a)}(1-v_{11}^{[11]})}\!\!+\!\rho^{\frac{4a}{3(1-a)}(1-v_{11}^{[12]})}\right)\!\!\nonumber\\
&~+\!\log(\rho^{2-v_{11}^{[11]}-v_{22}^{[11]}}\!\!\!+\!\rho^{2-v_{12}^{[11]}-v_{21}^{[11]}})\!\!<\!\log\rho^{\frac{r}{1-a}} \!\Big\}\nonumber\\
&\doteq\textrm{Pr}\Big\{\sum_{i=1,2}\{\rho^{\frac{4a}{3(1-a)}(1-v_{11}^{[1i]})+2-v_{11}^{[11]}-v_{22}^{[11]}}\nonumber\\
&~+\rho^{\frac{4a}{3(1-a)}(1-v_{11}^{[1i]})+2-v_{12}^{[11]}-v_{21}^{[11]}}\}<\rho^{\frac{r}{1-a}} \Big\},
\end{align}
where (f) follows from $\left(\rho^{1-v_{11}^{[11]}}+\rho^{1-v_{11}^{[12]}}\right)^{\frac{4a}{3(1-a)}}\geq \rho^{\frac{4a}{3(1-a)}(1-v_{11}^{[11]})}+\rho^{\frac{4a}{3(1-a)}(1-v_{11}^{[12]})}$.

Similar to proof of Theorem 1, we consider the two cases when $v_{11}^{[11]}<v_{11}^{[12]}$ and $v_{11}^{[11]}\geq v_{11}^{[12]}$. Omitting the details of calculation for economy of space, we have
$\textrm{Pr}\{\mathcal{E}_{IAA_{1}}\}
\doteq\rho^{-\min(\frac{2}{a+3},\frac{2}{3(1-a)}, \frac{1}{2a})(6-3r-2a)}$ when $v_{11}^{[11]}<v_{11}^{[12]}$, whereas $\textrm{Pr}\{\mathcal{E}_{IAA_{1}}\}
\doteq\rho^{-\min(\frac{2}{3(1-a)}, \frac{1}{2a})(6-3r-2a)}$ when $v_{11}^{[11]}\geq v_{11}^{[12]}$.
Combining the two cases, since $\frac{2}{a+3}<\frac{2}{3(1-a)}$ for $0 \leq a \leq 1$,
\begin{align}
\textrm{Pr}\{\mathcal{E}_{IAA_{1}}\}&\doteq\rho^{-\min\left(\frac{2}{a+3},\frac{2}{3(1-a)},\frac{1}{2a}\right)(6-3r-2a)}\nonumber\\
&\doteq\rho^{-\min\left(\frac{2}{a+3}, \frac{1}{2a}\right)(6-3r-2a)}.
\end{align}

For $\mathcal{E}_{IAA_2}$, the outage probability is given by
\begin{align}
\textrm{Pr}\{\mathcal{E}_{IAA_2}\}&=\textrm{Pr}\Big\{\!\sum_{i=1,2}\frac{4}{3}a\log\!\left(1\!+\!\rho\!\left(\left|h_{11}^{[i1]}\right|^2\!\!+\!\left|h_{11}^{[i2]}\right|^2\right)\right)\!\nonumber\\
&~+\!(1-a)\log\det\!\left(1+\rho\bar{\mathbf{H}}\bar{\mathbf{H}}^{\dagger}\right)
\!<\!2r\log\rho \Big\}\nonumber\\
&\doteq\textrm{Pr}\Big\{\log(\sum_{i,j=1,2}\rho^{2-v_{11}^{[1i]}-v_{11}^{[2j]}})^{\frac{4a}{3(1-a)}}\nonumber\\
&~+\log\!\!\!\!\!\sum\limits_{\scriptstyle l_1,l_2,k_1,k_2\in \{1,2\}\atop \scriptstyle\bigcap(l_1,k_1)\neq(l_2,k_2)}\rho^{2-v_{1k_1}^{[1l_1]}-v_{2k_2}^{[1l_2]}}<\log\rho^{\frac{2r}{1-a}} \Big\}\nonumber\\
&\leq\textrm{Pr}\Big\{\log\sum_{i,j=1,2}\rho^{(2-v_{11}^{[1i]}-v_{11}^{[2j]})\frac{4a}{3(1-a)}}\nonumber\\
&~+\log\!\!\!\!\!\sum\limits_{\scriptstyle l_1,l_2,k_1,k_2\in \{1,2\}\atop \scriptstyle\bigcap(l_1,k_1)\neq(l_2,k_2)}\rho^{2-v_{1k_1}^{[1l_1]}-v_{2k_2}^{[1l_2]}}<\log\rho^{\frac{2r}{1-a}} \Big\}\nonumber.
\end{align}
The dominant scale of $\textrm{Pr}\{\mathcal{E}_{IAA_2}\}$ is determine by the maximum scale among 48 possible combinations, similar to the on-off switched IA scheme based on the conventional IA.
If we consider a specific case, by the similar way of the on-off switched IA scheme based on the conventional IA case, the dominant scale of $\textrm{Pr}\{\mathcal{E}_{IAA_2}\}$ is obtained as
\begin{align}
\textrm{Pr}\{\mathcal{E}_{IAA_{2}}\}&\doteq \rho^{-\min(\frac{4}{3(1-a)},\frac{4}{a+3},\frac{1}{2a})(6-6r+2a)}\nonumber\\
&\doteq \rho^{-\min(\frac{4}{a+3},\frac{1}{2a})(6-6r+2a)}.
\end{align}
We skip the detailed proof of finding the dominant scale of $\textrm{Pr}\{\mathcal{E}_{IAA_2}\}$ for economy of page space.

Finally, the diversity gain of the on-off switched IA with Alamouti coding scheme is determined by the dominant scale of $\textrm{Pr}\{\mathcal{E}_{IAA_{1}}\}$ and $\textrm{Pr}\{\mathcal{E}_{IAA_{2}}\}$, which is also upper-bounded by the point-to-point 2 $\times$ 2 MIMO DMT.

\begin{IEEEbiography}
[{\includegraphics[width=1in,height =1.25in, clip,
keepaspectratio]{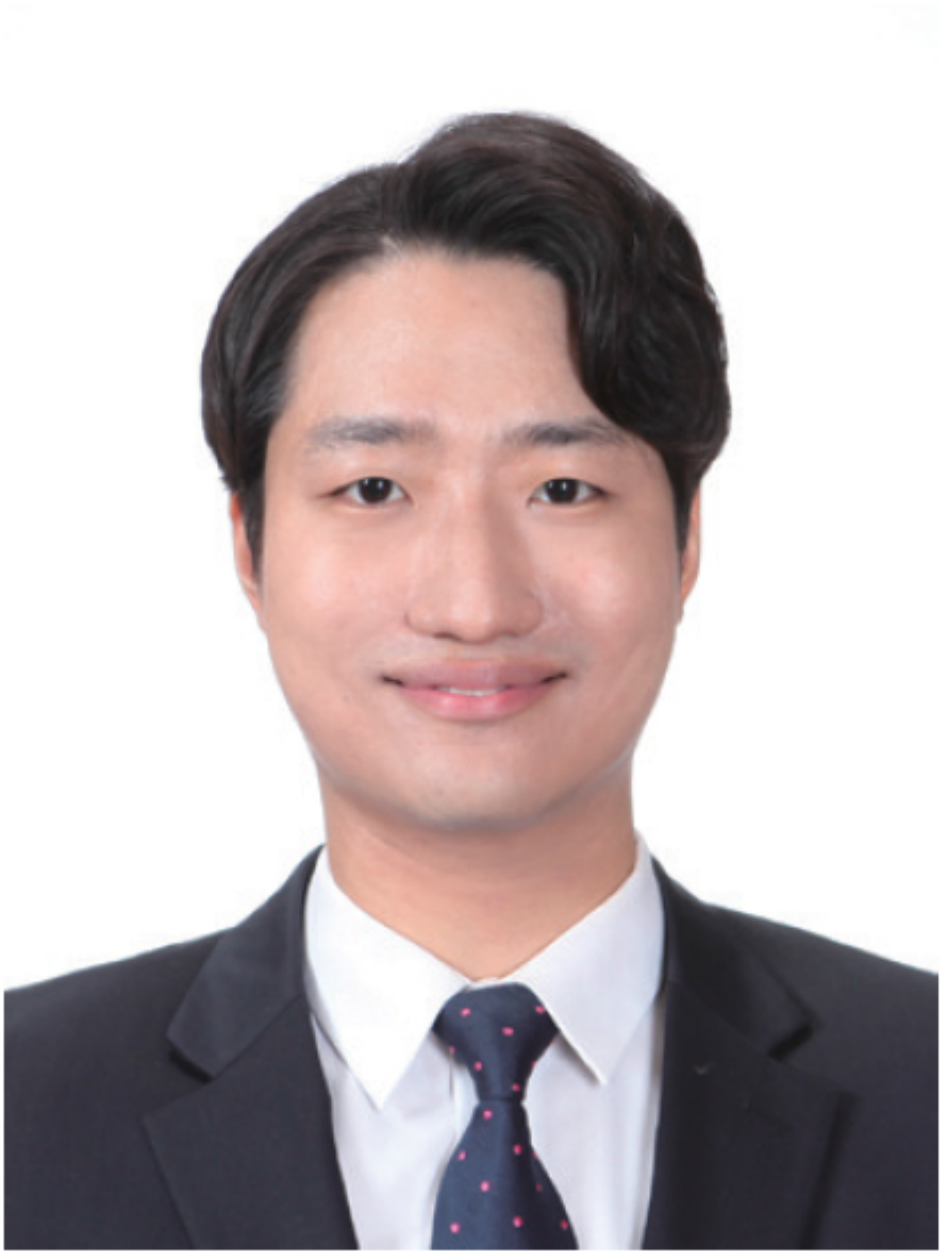}}] {Young-bin Kim} (S'11-M'17)
received the B.Sc. degree in Electrical Communications Engineering in 2008 and the M.Sc. and Ph.D. degree in Electrical Engineering from the Korea Advanced Institute of Science and Technology (KAIST), Daejeon, Korea, in 2010 and 2017, respectively. He was a Postdoctoral Research Scholar at KAIST from March to April in 2017. He is currently an Associate Research Engineer of KDDI Research, Inc., Saitama, Japan. 

He received the Korean Institute of Communications and Information Sciences (KICS) Outstanding Paper Award in 2016. He also received the IEICE 2018 Joint Conference on Satellite Communications Best Paper Award in 2018.
\end{IEEEbiography}
\begin{IEEEbiography}
[{\includegraphics[width=1in,height =1.25in, clip,
keepaspectratio]{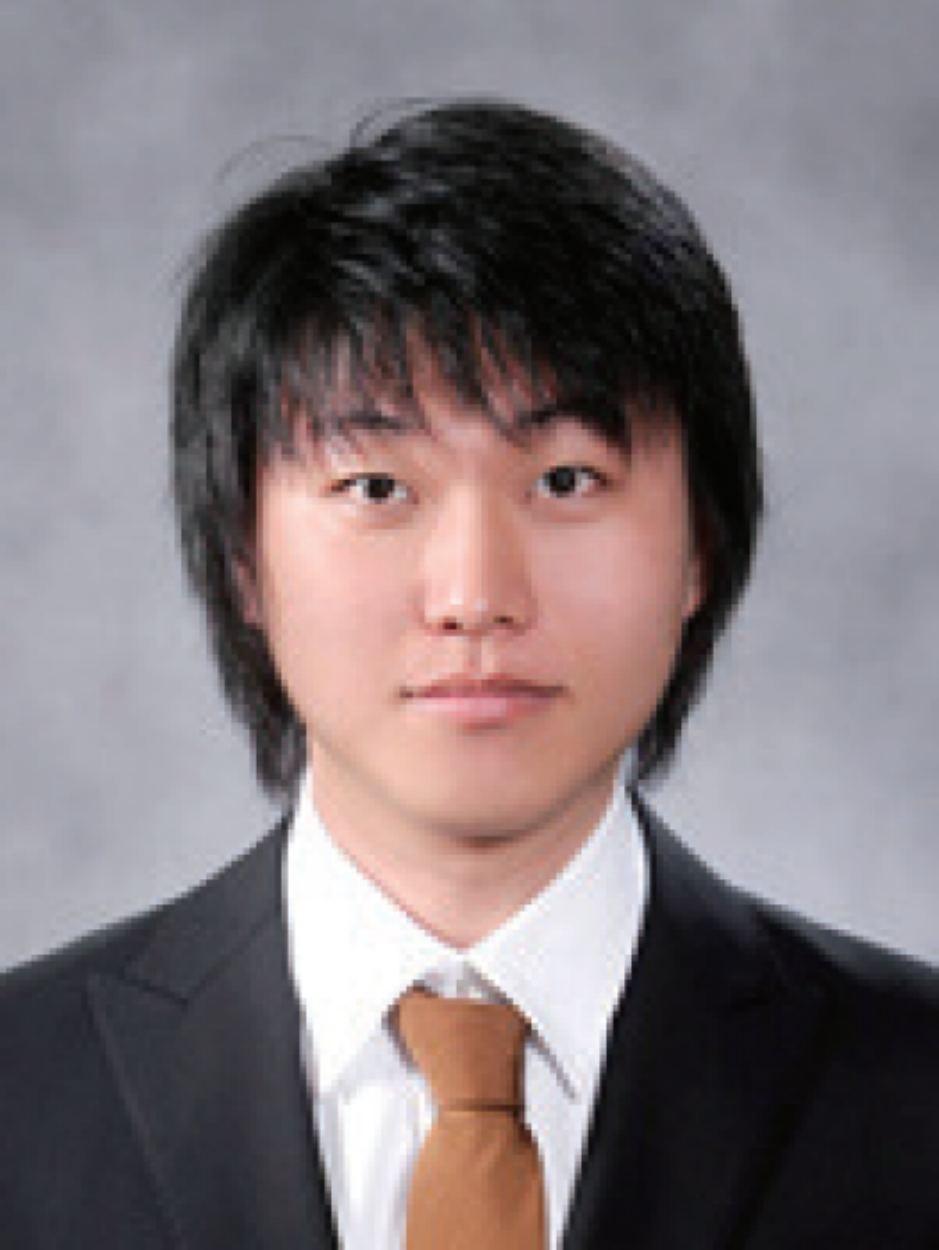}}] {Myung Gil Kang} (S'11-M'17)
received the B.S. degree in Information and Communications Engineering, M.S. degree in electrical engineering and the Ph.D. degree in electrical engineering from the Korea Advanced Institute of Science and Technology (KAIST), Daejeon, Korea, in 2009, 2011, and 2017, respectively. He was a Post-Doctoral Research Scholar at KAIST from March 2017 to February 2018 and a Post-Doctoral Research Scholar at Hankuk University of Foreign Studies, Yongin, Korea from April 2018 to July 2018. He is currently a Post-Doctoral Research Scholar at Mid Sweden University, Sundsvall, Sweden. His research interests include signal processing, interference management, wireless security and information theory.
\end{IEEEbiography}

\begin{IEEEbiography}
[{\includegraphics[width=1in,height
=1.25in,clip,keepaspectratio]{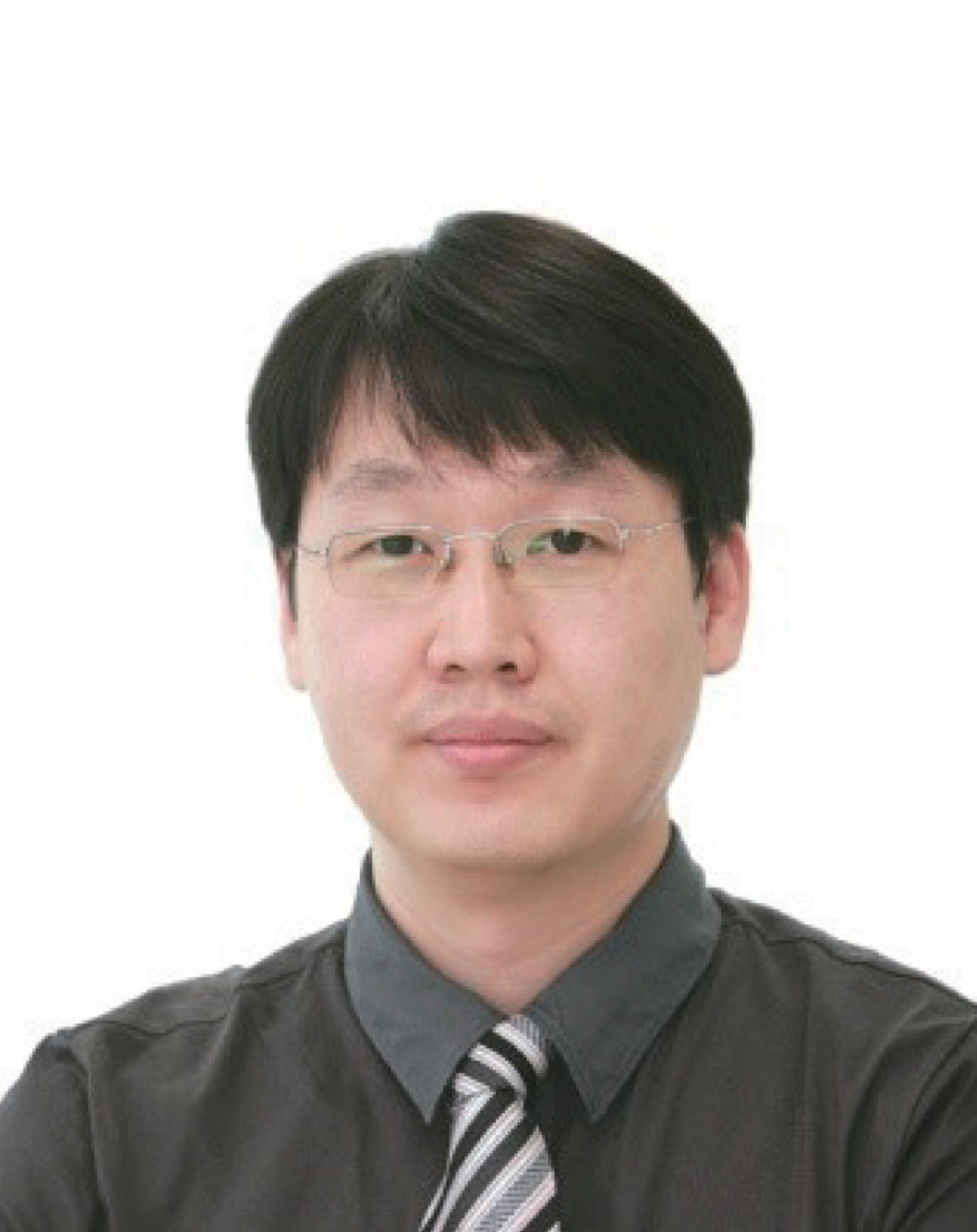}}] {Wan Choi}
(S'03-M'06-SM'12) received the B.Sc. and M.Sc. degrees from the School of Electrical Engineering and Computer Science (EECS), Seoul National University (SNU), Seoul, Korea, in 1996 and 1998, respectively, and the Ph.D. degree in the Department of Electrical and Computer Engineering at the University of Texas at Austin in 2006. He is currently Professor of School of Electrical Engineering, Korea Advanced Institute of Science and Technology (KAIST), Daejeon, Korea. From 1998 to 2003, he was a Senior Member of the Technical Staff of the R$\&$D Division of KT Freetel, Korea, where he researched 3G CDMA systems. 

He is the recipient of IEEE Vehicular Technology Society Jack Neubauer Memorial Award (Best System Paper Award) in 2002. He also received the IEEE Vehicular Technology Society Dan Noble Fellowship Award in 2006, the IEEE Communication Society Asia Pacific Young Researcher Award in 2007, and the Irwin-Jacobs Award from Qualcomm and KICS in 2015. While at the University of Texas at Austin, he was the recipient of William S. Livingston Graduate Fellowship and Information and Telecommunication Fellowship from Ministry of Information and Communication (MIC), Korea. He is an Executive Editor for the IEEE Transactions on Wireless Communications from Dec. 2014 and serves as Editor for the IEEE Transactions on Vehicular Technology. He also served as Editor for the IEEE Transactions on Wireless Communications (2009-2014), for the IEEE Wireless Communications Letter (2012-2017), and as Guest Editor for the 5G Wireless Communication System Special issue of the IEEE Journal on Selected Areas in Communications. 
\end{IEEEbiography}
\end{document}